\documentclass[article]{IEEEtran}
\usepackage{cite}

\usepackage{amsmath,amssymb,amsfonts}
\usepackage{mathtools}

\usepackage{algorithmic}
\usepackage{graphicx}
\usepackage{textcomp}
\usepackage{xcolor}
\usepackage{hyperref}
\usepackage{mathrsfs}
\usepackage{amsthm}
\usepackage{scalerel,stackengine}
\usepackage{paralist}
\usepackage{blkarray}
\usepackage[ruled,vlined,linesnumbered]{algorithm2e}
\usepackage{empheq}
\definecolor{lightgreen}{HTML}{90EE90}

\newenvironment{noinds_itemize}{\begin{list}{$\bullet$}
{\setlength{\rightmargin}{0em}
\setlength{\leftmargin}{1.2em}
\setlength{\itemsep}{0em}
\setlength{\topsep}{0em}
\setlength{\parsep}{0em}}}{\end{list}}

\def\descriptionnobflabel#1{\hspace\labelsep #1}
\def\descriptionnobf{\list{}{\labelwidth\z@ \itemindent-\leftmargin
 \let\makelabel\descriptionnobflabel}}

\def\s_descriptionlabel#1{\hspace\labelsep \bf #1}

\def\s_descriptionlabel#1{\hspace\labelsep \bf #1}

\newcommand{\bind}[1]{\hspace*{#1}\begin{minipage}[t]{7in}\begin{itemize}}
\newcommand{\eind}{\end{itemize}\end{minipage}\\}

\newcounter{ctr}

\newenvironment{s_enumerate-0}{\begin{list}{\thectr.}
{\usecounter{ctr}
\setlength{\rightmargin}{0cm}
\setlength{\leftmargin}{0cm}
\setlength{\itemsep}{0em}
\setlength{\topsep}{0em}
\setlength{\itemindent}{0cm}
\setlength{\parsep}{0em}}}{\end{list}}

\newenvironment{s_itemize-2}{\begin{list}{$\rhd$}
{\setlength{\rightmargin}{0em}
\setlength{\itemsep}{0em}
\setlength{\topsep}{0.25cm}
\setlength{\parsep}{0em}}}{\end{list}}

\newenvironment{itemize-1}{\begin{list}{$\bullet$}
{\setlength{\rightmargin}{0em}
\setlength{\itemsep}{0.5cm}
\setlength{\topsep}{0.25cm}
\setlength{\parsep}{0em}}}{\end{list}}

\newenvironment{itemize-2}{\begin{list}{$\rhd$}
{\setlength{\rightmargin}{0em}
\setlength{\itemsep}{0.25cm}
\setlength{\topsep}{0.25cm}
\setlength{\parsep}{0em}}}{\end{list}}

    {\end{list}}
    {\end{list}}

\newenvironment{itemize-3}{\begin{list}{-}
{\setlength{\rightmargin}{0em}
\setlength{\itemsep}{0.25cm}
\setlength{\topsep}{0.25cm}
\setlength{\parsep}{0em}}}{\end{list}}

\newenvironment{itemize-4}{\begin{list}{\ }
{\setlength{\rightmargin}{0em}
\setlength{\itemsep}{0.25cm}
\setlength{\topsep}{0.25cm}
\setlength{\parsep}{0em}}}{\end{list}}

\newenvironment{short-itemize}{\begin{list}{$\bullet$}
{\setlength{\rightmargin}{0em}
\setlength{\itemsep}{0.10cm}
\setlength{\topsep}{0.10cm}
\setlength{\parsep}{0em}}}{\end{list}}

\font\sf=cmss10
\newcommand{\Nats}{{\hbox{\sf I\kern-.13em\hbox{N}}}}   % Natural numbers
\newcommand{\Reals}{{\hbox{\sf I\kern-.14em\hbox{R}}}}  % Real numbers
\newcommand{\Ints}{{\hbox{\sf Z\kern-.43emZ}}}          % Integers
\newcommand{\CC}{{\hbox{\sf C\kern -.48emC}}}           % Complex numbers
\newcommand{\QQ}{{\hbox{\sf C\kern -.48emQ}}}           % Rational numbers

\newcommand{\lNats}{{\hbox{\sf {\large I\kern-.13em\hbox{N}}}}}   %Natural numbers
\newcommand{\lReals}{{\hbox{\sf {\large I\kern-.14em\hbox{R}}}}}  %Real numbers
\newcommand{\lInts}{{\hbox{\sf {\large Z\kern-.43emZ}}}}          %Integers
\newcommand{\lCC}{{\hbox{\sf {\large C\kern -.48emC}}}}           %Complex numbers
\newcommand{\lQQ}{{\hbox{\sf {\large C\kern -.48emQ}}}}           %Rational numbers

\usepackage[TS1,T1]{fontenc}
\usepackage[utf8]{inputenc}
\usepackage{lmodern}
\usepackage{nccmath}

\usepackage{newtxmath}
\usepackage{alltt}
\usepackage{epsfig}
\usepackage{verbatim}
\usepackage{fancybox}
\usepackage{subfigure}
\usepackage{url}
\usepackage{multirow}
\usepackage{setspace}
\usepackage{fancyhdr}
\usepackage{framed}
\usepackage{booktabs}
\usepackage{color}
\usepackage{wrapfig}
\usepackage{mathtools}%To number only eqns that are referred to
\mathtoolsset{showonlyrefs}% for this to work need to refer to eqns by \eqref, not \ref.
\usepackage{placeins}
\usepackage{tikz}

\newcommand\oast{\stackMath\mathbin{\stackinset{c}{0ex}{c}{0ex}{\ast}{\bigcirc}}}

\allowdisplaybreaks

\newtheorem{theorem}{Theorem}

\newtheorem{remark}{Remark}
\newtheorem{example}{Example}
\newtheorem{definition}{Definition}

\newtheorem{assumption}{Assumption}
\newtheorem{result}{Result}

\begin{document}

	\title{Graph Signal Processing: Dualizing GSP Sampling in the Vertex and Spectral Domains}
\author{John Shi, \IEEEmembership{Student Member, IEEE,} Jos\'e M.~F.~Moura, \IEEEmembership{Fellow, IEEE}%
\thanks{This material is based upon work partially funded and supported by the Department of Defense under Contract No. FA8702-15-D-0002 with Carnegie Mellon University for the operation of the Software Engineering Institute, a federally funded research and development center. This work is also partially supported by NSF grants CCF~1837607 and CCN~1513936.}%
\thanks{Department of Electrical and Computer Engineering, Carnegie Mellon University, Pittsburgh PA 15217 USA; [jshi3,moura]@andrew.cmu.edu.}}
	
	\maketitle
	
	\begin{abstract}
		%Discrete Signal Processing (DSP) studies well-ordered time and image signals.
		Vertex based and spectral based GSP sampling has been studied recently. The literature recognizes that methods in one domain do not have a counterpart in the other domain. This paper shows that in fact one can develop a unified graph signal sampling theory with analogous interpretations in both domains just like sampling in traditional DSP. To achieve it, we introduce a spectral shift~$M$ acting in the spectral domain rather than shift~$A$ that acts in the vertex domain. This leads to a GSP theory that starts from the spectral domain, for example, linear shift invariant (LSI) filtering in the spectral domain is with polynomial filters $P(M)$. We then develop GSP vertex and spectral domain dual versions for each of the four standard sampling steps of subsampling, decimation, upsampling, and interpolation. We show how GSP sampling reduces to DSP sampling when the graph is the directed time cycle graph. Simple examples illustrate  the impact of choices that are available in GSP sampling.
		\end{abstract}
	\vspace*{-.4cm}
	\section{Introduction}\label{sec:introduction}
	%%%%%%%%%%%%%%%%%%%%%%%%%%%%%%%%%%%%%%%%%%%%%%%%%%%%%%%%%%
	%%%%%%%%%%%%%%%%%%%%%%%%%%%%%%%%%%%%%%%%%%%%%%%%%%%%%%%%%%
%%%%%%%%%%%%%%%%%%%%%%%%%%%%%%%%%%%%%%%%%%%%%%%%%%%%%%%%%%
%%%%%%%New Introduction
%\subsection{New Introduction}
With time signals, samples are referenced to or indexed by time, while with images by pixels. Time takes value in a subset of the integers $\mathbb{Z}$ and pixels in a subset of the Cartesian product $\mathbb{Z}\times \mathbb{Z}$. These are \textit{regular} indexing structures. With modern applications, data is often indexed by irregular structures like graphs. Graph Signal Processing (GSP) \cite{Sandryhaila:13,ShumanNFOV:13,Sandryhaila:14,Sandryhaila:14big} has been developed in the last decade to study graph (indexed) data. GSP as in \cite{Sandryhaila:13,Sandryhaila:14,Sandryhaila:14big} begins with a graph interpretation of Discrete Signal Processing~(DSP) where the signal time samples become data indexed by the nodes of a directed cycle graph.\footnote{\label{ftn:timesignals1} For simplicity, the paper assumes finite discrete time signals that are either periodic or have periodic extensions.} This interpretation is then extended \cite{Sandryhaila:13,Sandryhaila:14,Sandryhaila:14big} to data indexed by the nodes of arbitrary \textit{directed} or \textit{undirected} graphs $G=(V,E)$ defined by $N\times N$ adjacency matrix~$A$. These references \cite{Sandryhaila:13,Sandryhaila:14,Sandryhaila:14big} extend several DSP concepts to GSP, including graph shift, graph filtering, graph Fourier transform~(GFT), graph frequency, and graph filter response. In GSP, the adjacency matrix $A$ becomes the \textit{shift} operator, and it plays in GSP the same role that the time shift $z^{-1}$ plays in DSP. Reference \cite{ShumanNFOV:13} presents an alternative ``spectral domain'' development of GSP that is restricted to data indexed by nodes of \textit{undirected} graphs. It starts from spectral decompositions of the data in terms of the eigenfunctions of a variational operator, the graph Laplacian~$L$. For \textit{undirected} graphs, $L$ and~$A$ have the same eigendecomposition and, for undirected graphs, the spectral analysis is equivalent for the two approaches.

The approach in \cite{Sandryhaila:13,Sandryhaila:14,Sandryhaila:14big} develops GSP to parallel DSP as much as possible. For example, in DSP linear shift invariant (LSI) filters are finite degree polynomials of the shift $z^{-1}$. Likewise, LSI filters in GSP are matrix polynomials in~$A$ \cite{Sandryhaila:13,Sandryhaila:14,Sandryhaila:14big}. This approach to GSP is intuitively pleasing:
\begin{inparaenum}[1)]
\item new concepts in GSP are often designed as natural extensions of DSP concepts;
    \item GSP is a generalization of DSP, i.e., when the underlying graph~$G$ is a cyclic graph, GSP becomes DSP; and
        \item GSP leads to reinterpretation of well known DSP results and sheds new light on DSP facts that are commonly taken for granted.
            \end{inparaenum}
These points will become apparent in this paper.

Recently, \cite{EldarTanakaSPM} reviews comprehensively the existing methods for graph sampling. According to \cite{EldarTanakaSPM}, ``two definitions [of graph sampling] can be possible $\cdots$'' and current theories are either in the vertex or in the graph spectral domain, with no simple analogy between them. This stands in contrast with the \textit{dualism} between time and frequency domain approaches to sampling. The reference identifies the ``[I]nterconnection between vertex and spectral representations of sampling ...'' as an open issue worthy of further study. It further asks how ``can these sampling approaches be described in a more unified way beyond a few known special cases?'' stating that ``This may lead to a more intuitive understanding of graph signal sampling.'' Our paper deals with this open issue.

\begin{figure}[hbt]
    \centering
   \includegraphics[width=8cm,keepaspectratio]{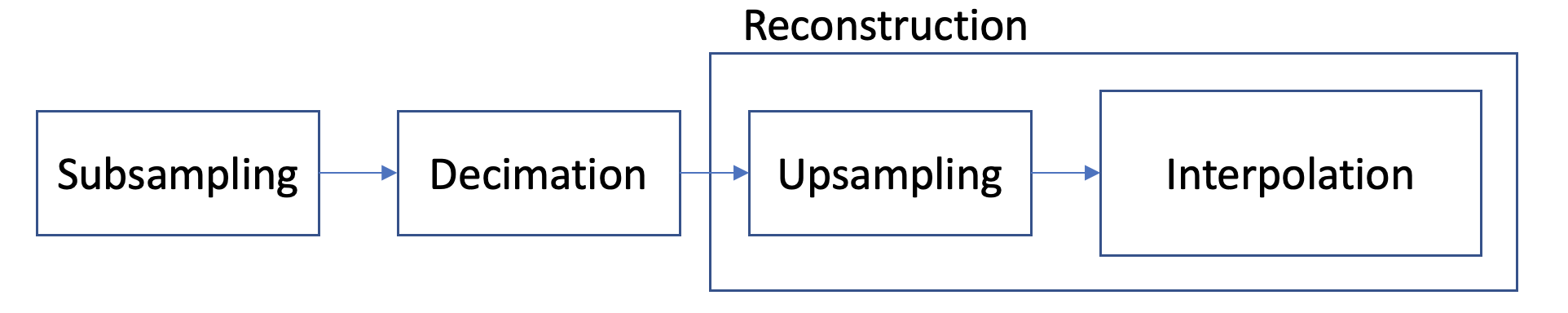}
    \caption{\small The sampling and reconstruction steps in this paper. Each step is considered in both vertex and spectral domains.}
    \label{fig:paths}
\end{figure}

\noindent\textbf{Paper contributions.} Below, we discuss how to pick a sampling signal or sampling set, but we do not claim it to be original. Other available methods in the literature can be used. We summarize our contributions.
\begin{noinds_itemize}
%\begin{s_itemize_3}
\item\textit{GSP sampling\textemdash the dualism between vertex and spectral representations of sampling.}\label{sit:GSPdual1}
Our presentation of graph sampling parallels the traditional sampling of discrete time signals \cite{oppenheimwillsky-1983}. The paper shows for graph sampling
\begin{inparaenum}[i)]
\item how it is analogous to the Shannon-Nyquist and shift-invariant sampling of time signals; and
    \item how and when it deviates given the intrinsic differences between GSP and DSP constructs.
         \end{inparaenum}
         Safeguarding the distinctions, we replicate the dualism between time and frequency sampling operations in DSP with a similar dualism between GSP sampling operations in the vertex and the graph spectral domains. For every sampling operation, we present its vertex domain interpretation \textit{and} its spectral domain interpretation. When GSP becomes DSP, we point out which among several alternative choices of the sampling set leads to the Shannon-Nyquist uniform sampling of time signals.

         \phantom{Th}We detail our contributions to the GSP sampling and reconstruction steps: subsampling, decimation, upsampling, and interpolation (shown in figure \ref{fig:paths}).
\item \textit{Graph subsampling: LSI spectral filtering.}\label{sit:GSPdual1=2} Subsampling of a graph signal is multiplication in the vertex domain of the graph signal~$s$ by a zero-one graph sampling signal $\delta^{\scriptsize\textrm{spl}}$. To determine the graph spectral domain operation that is dual to multiplication of graph signals in the vertex domain, we introduce graph LSI filtering in the graph spectral domain. Namely, we introduce graph convolution in the spectral domain by a polynomial spectral filter $P(M)$ using a new graph spectral shift~$M$ that we define. This dualizes multiplication of graph signals in the vertex domain with graph convolution or graph filtering in the spectral domain. Further, spectral shift~$M$ defines a new spectral graph whose nodes now index the graph Fourier coefficients by the graph frequencies. The interesting point is that subsampling by~$\delta^{\scriptsize\textrm{spl}}$ is in the spectral domain achieved by LSI spectral filtering (by $P(M)$) just like in DSP sampling.
    \item \textit{Graph subsampling: Spectral replication.}\label{sit:GSPdual1=3a} Like in DSP, in GSP we show that the spectrum of the subsampled signal is a set of (distorted) replicas of the lowpass spectrum of the original signal, with no aliasing if sampling rate is at least the ``GSP Nyquist'' rate.
    \item\textit{Graph decimation.} Decimation downsizes the original graph from a graph of order~$N$ to a graph of order~$K$. This is determined by
         \begin{inparaenum}[i)]
         \item the choice of sampling signal (and sampling set), and
         \item by the~$K$ nonzero components of the graph spectrum of the signal.
         \end{inparaenum}
         This follows standard procedure, for example \cite{chenvarmasinghkovacevic}. Just like for DSP, this step expands the spectrum of the decimated graph signal to the full band (which is now reduced to a smaller number of graph frequencies). But, even though where the graph spectrum is nonzero is given or assumed, there are several choices for the sampling signal, and so for the decimated graph. We illustrate why and what choices lead to the uniform sampling in DSP. In DSP, certain choices preserve the nature of the graph, i.e., both the original and the decimated graph are cycle graphs (of different orders), while other choices lead to a decimated graph that is no longer cyclic. In GSP, this is hardly the case. For general graphs, the decimated and original graphs are very different.
    \item \textit{Reconstruction: upsampling.} The upsampling with reinsertion of zeros in the decimated graph signal, and the reconstruction of the original $N$th-order graph follows the same steps in GSP and DSP. Like in DSP, it leads to a contracted signal spectrum in GSP, i.e., a signal spectrum that is zero in~$N-K$ components.
        \item \textit{Reconstruction: interpolation by spectral filtering and ideal LSI (vertex) filtering.} Perfect reconstruction in GSP sampling, assuming no aliasing, is obtained like in DSP by spectral filtering, but in two distinct steps. First, with a spectral filter~$Q$ that is not necessarily LSI, followed by LSI vertex ideal filtering. When the graph is cyclic and GSP becomes DSP, the reconstruction filter~$Q$ becomes trivially a gain of~$K$, and the LSI vertex ideal sampling is the Shannon ideal filter with a sinc function as impulse response (in the vertex domain).
\end{noinds_itemize}

%%%%%%%%%%%%%%%%%%%%%%%%%%%%%%%%%%%%%%%%%%%%%%%%%%%%%%%%%%
%%%%%%%%%%%%%%%%%%%%%%%%%%%%%%%%%%%%%%%%%%%%%%%%%%%%%%%%%%
	%%%%%%%%%%%%%%%%%%%%%%%%%%%%%%%%%%%%%%%%%%%%%%%%%%%%%%%%%%
	%
		\textbf{Brief review of the literature on GSP sampling}. In the last decade, there has been a robust literature on GSP sampling. With a few exceptions, these works assume \textit{undirected} graphs. Due to space limitations, we  highlight a few main points and refer to the references and their bibliography for a more comprehensive review. Most of the work has been concerned with choosing the sampling set and developing recovery methods that address issues like noise, or aliasing, or robustness to computational errors, or speed of computation. These are important issues, but are not our focus. Our paper is concerned with what is lacking \cite{EldarTanakaSPM}, namely, presenting a unified GSP sampling theory and the analogy (duality) between vertex and spectral GSP sampling operations, just like for DSP sampling. Still, a brief review of the literature helps put our work in context.
	
	References~\cite{pesenson2008sampling,pesenson2010sampling} consider subspaces of bandlimited signals (Paley-Wiener spaces) and show that signals supported by undirected graphs can be perfectly reconstructed from values in a \textit{sampling} set~$S$, termed \textit{uniqueness} set. Critically sampled graph signals \textit{restricted to}  undirected $k$-regular bipartite graphs are considered in \cite{Narang:12} that proposes two-channel wavelet filter banks for perfect reconstruction. General undirected graphs are approximated by a decomposition in terms of $k$-regular bipartite graphs.
	%results of  \cite{harary1977biparticity}.
	Papers  \cite{anis2014towards,gadde2015probabilistic,anis2016efficient,anis2017critical} consider choosing~$S$ for stable reconstruction and methods that avoid spectral decompositions of the graph Laplacian. While the previous references apply to undirected graphs, \cite{Jelena} shows  that a necessary and sufficient condition for perfect reconstruction is for the sampling set to choose~$K$ linearly independent rows from~$K$ columns of the inverse graph Fourier transform. The paper shows that random sampling chooses with high probability for Erd\"os R\'enyi graphs a sampling set from which perfect reconstruction can be achieved. This reference also shows that sampling preserves first order differences on the sampled nodes. In \cite{chenvarmasinghkovacevic}, various sampling schemes are considered including uniform sampling, experimentally designed sampling, and active sampling. Random sampling for undirected graphs is also studied in \cite{tremblay2017graph} that presents a condition in terms of the invertibility of a $K\times K$ \textit{kernel} matrix corresponding to rows and columns selected by a determinantal point process. Reference \cite{tsitsvero2016signals} derives an uncertainty principle for graph signals and conditions for  recovery of bandlimited signals from a subset of the samples. In \cite{marques2015sampling},  the authors propose a sampling scheme that uses as input observations taken at a single node and corresponds to sequential applications of the graph-shift operator. Beyond sampling of bandlimited signals, \cite{jungheromarajahromiheimowitzeldar-2019} considers piecewise constant signals on undirected graphs and \cite{chen2015signal-2} smooth signals, with recovery by a Lasso like procedure. Reference \cite{eldar-samplingbook2015} reconstructs time signals from projections on low rank approximation subspaces, with \cite{gensample} extending it to  undirected graph signals. Reference \cite{tanaka-2018}  presents a spectral domain sampling where the spectrum of the subsampled signal replicates the bandlimited spectrum of the original signal. But the method in \cite{tanaka-2018} does not respect the traditional concept of sampling, namely, discarding samples in the vertex domain and reconstructing the original signal from the samples kept. In fact, the sampled signal in \cite{tanaka-2018} not only keeps all samples of the original signal, but also distorts them.

Our paper does not claim novelty in finding the sampling set~$S$. We assume that~$S$ is given. In DSP, this is similar to assuming a particular sampling scheme, say, uniform sampling (keeping every $K$th sample). In DSP, the following \textit{duality} holds:
\begin{inparaenum}[1)]
\item subsampling in the frequency domain is linear shift invariant filtering leading to spectral replication, and
    \item reconstruction is by ideal lowpass filtering.
     \end{inparaenum}
     Likewise, extending the DSP sampling framework to GSP sampling in a natural way, we show similar dualism:
     \begin{inparaenum}[1)]
     \item GSP subsampling in the vertex domain is LSI filtering in the spectral domain, leading to spectral replication, and
     \item when is GSP reconstruction achieved by LSI filtering.
     \end{inparaenum}
     To develop this, we introduce a spectral shift~$M$, a spectral graph $G_{\scriptsize\textrm{sp}}$, spectral GSP filtering, and other concepts, and we show which choices among alternatives in GSP replicate Shannon-Nyquist sampling in DSP.

\textbf{Guide to paper}. Section~\ref{sec:primerongsp} reviews basic GSP concepts.  Section~\ref{sec:spectralshift} introduces spectral shift~$M$, its properties, and $\textrm{GSP}_{\textrm{sp}}$, a dual to GSP from the spectral domain point of view, derived from~$M$. Section~\ref{sec:subsamplingdecimation} considers GSP subsampling and GSP decimation in both vertex and spectral domains. Section~\ref{sec:upsamplinginterpolation} considers GSP upsampling and interpolation. Finally, section~\ref{sec:conclusion} concludes the paper.
%Section~\ref{sec:graphsignalrep} considers graph signal representations for GSP and $\textrm{GSP}_{\textrm{sp}}$, introducing alternative definitions for graph impulse, various signal representations including graph filter representations $P_s(A)$ and $P_s(M)$ of signal~$s$, and polynomial representations. Section~\ref{sec:graphconvolution} studies graph convolution in the vertex and spectral domains and for different signal representations, introducing a \textit{fast GSP convolution} that uses the FFT, i.e., the DSP FFT algorithm. Section~\ref{sec:gspsampling} considers GSP sampling from solution of linear systems relating spectral and vertex domain signal representations, discussing sampling set selection, necessary and sufficient conditions for perfect reconstruction, and then subsampling as LSI spectral filtering and signal interpolation or reconstruction as spectral filtering. The section also relates the approach of this paper to the spectral sampling approach in \cite{tanaka-2018,gensample}. Finally, section~\ref{sec:conclusion} concludes the paper.

 %
\section{Primer on GSP}\label{sec:primerongsp}
\begin{figure}[h!]
%\vspace*{-.5cm}
%\centering	\includegraphics[height=2.25cm,width = 4cm]{images/ring.png}
\centering	\includegraphics[height=1.cm,width = 3in]{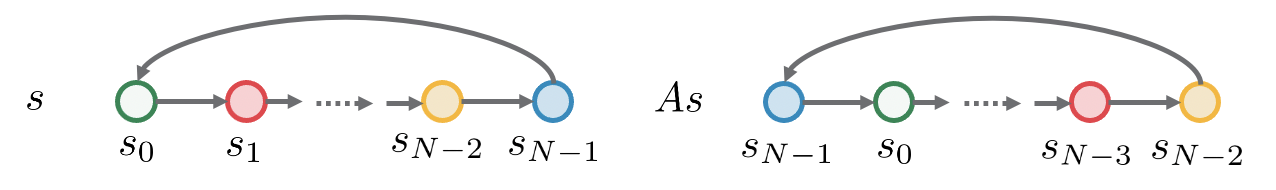}
	\vspace*{-.4cm}\caption{Ring graph: Left: graph signal~$s$; right: its shifted $A\cdot s$.}
		\label{fig1}
	\end{figure}

%
% \vspace*{-.5cm}
%
	We provide a brief review of GSP following \cite{Sandryhaila:13,Sandryhaila:14big}, see also \cite{ortegafrossardkovacevicmouravandergheynst-2018}. As a motivation, we start by casting DSP of time signals of finite length~$N$  (equivalently, periodic time signals) in the context of GSP. We collect the~$N$ time signal samples $s_0,s_1,\cdots, s_{N-1}$ in the vector $s=\left[s_0\,s_1\,\hdots\,s_{N-1}\right]^T$. Consider a cycle graph of~$N$ nodes and let each signal sample $s_n$ be indexed by a node of the graph  (left of Figure~\ref{fig1}).
	\vspace{-.2cm}
For this graph, the adjacency matrix is cyclic
%{\small
\begin{align}
\label{eqn:graphshiftA-1}
	A &{}= \left[\begin{array}{ccccc}
	0  & 0  & \hdots & 0 & 1 \\
	1  & 0  &  \hdots & 0 & 0 \\
	%0  & 1 & 0 &  \hdots & 0 & 0 \\
	\vdots  & 1 & \ddots & \vdots  & \vdots\\
	\vdots  & \vdots & \ddots & \ddots  & \vdots\\
	0  & 0 &  \hdots & 1 & 0 \\
	\end{array}\right].
	\end{align}%%
%}%
	%
	
In GSP, the adjacency matrix~$A$ in~\eqref{eqn:graphshiftA-1} is also the matrix representation of the DSP shift $z^{-1}$\textemdash we shift the signal~$s$ by multiplication with the shift~$A$ to get the shifted signal $A\cdot s$, shown on the right of Figure~\ref{fig1}.

	The Fourier transform for finite time signals is the Discrete Fourier Transform~(DFT). It can be found through the eigendecomposition of the shift matrix~$A$:
%{\small
	\begin{align}\label{Adecomp}
	A &{}= \textrm{DFT}^\textrm{\,-1} \cdot \Lambda \cdot \textrm{DFT}=\textrm{DFT}^\textrm{H} \cdot \Lambda \cdot \textrm{DFT}.	\end{align}
In~\eqref{Adecomp}, $\Lambda$ is the diagonal matrix of eigenvalues of~$A$:
\begin{align}
\label{eqn:dspLambda-2}
\Lambda{}&=\left[\begin{array}{ccc}
\lambda_0&&\\
&\ddots&\\
&&\lambda_{N-1}
\end{array}\right], \,  \lambda_k=e^{-j\frac{2\pi}{N}k},\: k=0,\cdots, (N-1),\\
{}&=\textrm{diag}\left[\lambda_0,\cdots,\lambda_{N-1}\right],
\label{eqn:dspLambda}
\end{align}
%}%
and the $\textrm{DFT}$ is the discrete Fourier transform matrix
\begin{align}
\label{eqn:DFT1}
\hspace{-.5cm}\textrm{DFT}{}&=\frac{1}{\sqrt{N}}\left[\begin{array}{llll}
1&1&\cdots&1\\
1&e^{-j\frac{2\pi}{N}}&\cdots&e^{-j\frac{2\pi}{N}(N-1)}\\
\vdots&\vdots&&\vdots\\
1&e^{-j\frac{2\pi}{N}(N-1)}&\cdots&e^{-j\frac{2\pi}{N}(N-1)(N-1)}
\end{array}\right]\!.
\end{align}
By~\eqref{eqn:DFT1}, the $\textrm{DFT}$ is symmetric and unitary, so $\textrm{DFT}^H=\textrm{DFT}^*=\textrm{DFT}^{-1}$ as used in~\eqref{Adecomp}.

The spectrum of~$A$, i.e., its eigenvalues $\left\{e^{-j\frac{2\pi}{N}k}\right\}_{0\leq k\leq N-1}$, are the frequencies.\footnote{\label{ftn:DSPfrequencies}In DSP, it is also common to refer to $\Omega_k=\frac{k}{N}=-\frac{1}{2\pi j}\ln e^{-j\frac{2\pi}{N}k}$ as the frequencies rather than to the eigenvalues $\lambda_k$.} The columns
\begin{align}\label{eqn:keigenvectorA}
v_k{}&=\frac{1}{\sqrt{N}}\left[1 \,\,\,e^{j\frac{2\pi}{N}k}\,\,\cdots\,\, e^{j\frac{2\pi}{N}k(N-1)}\right]^T, \,\,\,k=0,\cdots, N-1.
\end{align}
of $\textrm{DFT}^H$ are the eigenvectors of~$A$. They are the spectral components or harmonics of time signals.
  \begin{remark}\label{rem:notation1}
       The subindex~$k$ may refer to the $k$th vector or the $k$th entry of a vector. The context should remove the ambiguity. Our convention is to start indices from~$0$.
      \end{remark}
\textbf{Graph Shift}. GSP \cite{Sandryhaila:13,Sandryhaila:14big} extends DSP to indexing sets~$V$ that are the vertex sets of arbitrary directed or undirected graphs $G=(V,E)$. The graph signal~$s\in\mathbb{C}^N$ assigns a data sample~$s_n$ to vertex or node~$n\in V$, $n=0\cdots N-1$. Following \cite{Sandryhaila:13}, the graph shift is the adjacency matrix\footnote{Other authors consider other shifts, e.g., the symmetric and positive semi-definite graph Laplacian \cite{ShumanNFOV:13} that is restricted to undirected graphs, or unitary variations of~$A$ that sacrifice locality \cite{giraultgoncalvesfleury-2015,gavilizhang-2017}.} $A$. We shift~$s$ by applying shift~$A$ to~$s$, i.e., $A\cdot s$.

In this paper, unless otherwise stated, we assume:
\begin{assumption}[Diagonalizability]\label{assp:diagonalizableA} $A$ is diagonalizable.
\end{assumption}
Diagonalizability holds if~$A$ is symmetric or has distinct eigenvalues. We explicitly state when we assume the latter.
\begin{assumption}[Distinct eigenvalues]\label{assp:distincteigenA} Shift~$A$ has distinct eigenvalues.
\end{assumption}
\textbf{Graph Fourier Transform (GFT)}. Let the eigendecomposition of arbitrary shift~$A$ be
	\begin{align}\label{Adecompgsp}
	A {}&= \textrm{GFT}^\textrm{-1} \cdot \Lambda \cdot \textrm{GFT}.
	\end{align}
where $\Lambda$ is the diagonal matrix of the eigenvalues $\left\{\lambda_k\right\}_{0\leq k\leq N-1}$ of~$A$ and $\textrm{GFT}^\textrm{-1}$ is the matrix of its eigenvectors $\left\{v_k\right\}_{0\leq k\leq N-1}$. The graph Fourier Transform~(GFT) is the matrix $\textrm{GFT}$ in~\eqref{Adecompgsp}.\footnote{\label{ftn:GSPfrequencies} If assumption~\ref{assp:diagonalizableA} does not hold, the shift is non diagonalizable, see then \cite{Sandryhaila:13,derimoura-2017} for the definition of the $\textrm{GFT}$.}
	The eigenvalues $\lambda_k$ are the graph frequencies and the eigenvectors are the graph spectral components (like the harmonics for time signals).

\begin{remark}[Uniqueness: $\textrm{GFT}$ and~$\Lambda$]\label{rmk:uniquenessGFT-1} The eigendecomposition of a matrix is defined up to permutation of its eigenvalues and normalization of its eigenvectors. We assume that, when referring to the eigendecomposition~\eqref{Adecompgsp}, this ordering and normalization have been fixed, so~$\Lambda$, $\textrm{GFT}$, and $\textrm{GFT}^{-1}$ have been uniquely defined by~\eqref{Adecompgsp}.
\end{remark}
	
%	To refer to the columns and rows of both matrices, let:
%	\begin{align}\label{eqn:GFTinvcolumnrow}
%	   \textrm{GFT}^{-1}{}&=\left[v_0\cdots v_{N-1}\right]
%	   =\left[\begin{array}{c}
%	   u_0^H\\
%	   \cdots\\
%	   u_{N-1}^H
%	   \end{array}\right]
%	   \\
%	   \label{eqn:GFTcolumnrow}
%	    \textrm{GFT}\,\,\,\,\,{}&=\left[y_0\cdots y_{N-1}\right]
%	   =\left[\begin{array}{c}
%	   w_0^H\\
%	   \cdots\\
%	   w_{N-1}^H
%	   \end{array}\hspace{-.09cm}\right].
%	\end{align}
%	The Fourier basis are the columns $v_0,\cdots, v_{N-1}$ of $\textrm{GFT}^{-1}$.
	
    \textbf{Graph impulse}. In DSP, the impulse $\delta_0=[1\,0\cdots 0]^T\xrightarrow{\mathcal{F}}\widehat{\delta}_0=\frac{1}{\sqrt{N}}1$. In GSP, we have a choice when defining the graph impulse\textemdash either make it impulsive in the vertex domain or flat in the spectral domain. The first choice ties the graph impulse to picking the vertex where it is nonzero, while the second forces no such choice (constant across all frequencies). Let~$1$ be the column vector of all ones. We define the graph impulse through its $\textrm{GFT}$ as
    \begin{align}\label{eqn:deltaflat-1}
    \delta_0^{\scriptsize\textrm{flat}}{}&\xrightarrow{\mathcal{F}}\widehat{\delta}_0^{\scriptsize\textrm{flat}}=\frac{1}{\sqrt{N}}1.
    \end{align}

	\textbf{Filtering in the vertex domain}. In GSP, linear, shift invariant (LSI) filters are polynomials $P(A)$ of the shift~$A$,
 \begin{align}\label{eqn:convtime-11a}
	P(A) {}&=p_0I+p_1A+\cdots+p_{N-1}A^{N-1},
	\end{align}
and LSI graph \textit{filtering} is matrix-vector multiplication  \cite{Sandryhaila:13}
 \begin{align}\label{eqn:convtime}
	t{}&=P(A) \cdot s=\left[p_0I+p_1A+\cdots+p_{N-1}A^{N-1}\right]\cdot s.
	\end{align}

  \textbf{Graph frequency response}. For graph filter $P(A)$,
 \begin{align}\label{eqn:P(A)-1}
 P(A){}&=\textrm{GFT}^{-1}P(\Lambda)\,\textrm{GFT}\\
\label{eqn:P(Lambda)-1}
  \hspace{-1cm}\textrm{with  }\hspace{1.5cm}
P(\Lambda){}&=\textrm{diag}\left[P\left(\lambda_0\right),\,\cdots,\, P\left(\lambda_{N-1}\right)\right],
\end{align}
    where $P\left(\lambda_n\right)$ is $P(A)$ evaluated at the eigenfrequency $\lambda_n$.

     The \textit{graph frequency response} $p(\lambda)$ of $P(A)$ is 
 \begin{align}\label{eqn:graphfrequencyresponse-1a}
 p(\lambda){}&=P(\Lambda)\cdot\frac{1}{\sqrt{N}}1.
 \end{align}
 \begin{remark}\label{rmk:redefinep(lambda)}
 In the sequel, we absorb the $\frac{1}{\sqrt{N}}$ in the polynomial coefficients
  \begin{align}\label{eqn:p(lambda)-1}
p(\lambda){}&=P(\Lambda)\cdot 1=\left[\begin{array}{ccc}
P\left(\lambda_0\right)&
\cdots&
P\left(\lambda_{N-1}\right)
\end{array}\right]^T.
 \end{align}
 \end{remark}

 \textbf{Filtering in the frequency domain}.
 Filtering in the spectral domain then becomes:
 	\begin{align}
 	\label{eqn:convtime-2}
	\widehat{t}&{}=P(\Lambda) \cdot \widehat{s}
%\label{eqn:convtime-3}
%	&{}
	=(P(\Lambda)\cdot 1) \odot \widehat{s}=p(\lambda)\odot \widehat{s}
	\end{align}
 	where~$\odot$ is the Hadamard or pointwise or componentwise product of the frequency response $p(\lambda)$ and $\widehat{s}$.
  %$P\left(\lambda_k\right)$  %see Notation on page \pageref{pgr:p(lambda)},

 %
%
  Equations~\eqref{eqn:convtime} and~\eqref{eqn:convtime-2} are the equivalent versions of filtering in the vertex and spectral domains in GSP: graph filtering in the vertex domain multiplies the vector signal~$s$ by the matrix filter $P(A)$. In the spectral domain, it is the product of the \textit{diagonal} matrix filter $P(\Lambda)$ with the graph Fourier transformed~$\widehat{s}$, or, equivalently, it is the pointwise product~$\odot$ of the graph frequency response $p(\lambda)$ with~$\widehat{s}$.

 	%References \cite{Sandryhaila:13,Sandryhaila:14,Sandryhaila:14big, vandermod} provide fundamental GSP operations, they do not define shift in the spectral domain, convolution in the spectral domain, or multiplication in the spectral domain. Furthermore, while they define convolution in the vertex domain through filtering (matrix (filter) vector (graph signal) product), they do not define convolution of two graph signals, nor find the graph filter $P(A)$ given its graph impulse response. We will discuss relations to a prior work \cite{vandermod} in Section \ref{sec:modconvolution}. In order to consider sampling in both domains, we need to define these operations. To address these issues, we consider shifting in the graph spectral domain first.
%\vspace*{-.3cm}
\textbf{LSI filter $P(A)$ and its frequency response $\widehat{h}$}. Given a polynomial filter $P(A)$ with coefficients $p=\left[p_0\hdots p_{N-1}\right]$, equation~\eqref{eqn:p(lambda)-1} gives the frequency response $p(\lambda)$ of $P(A)$. The next result is the reverse: given a frequency response~$\widehat{h}$, determine the filter $P_h(A)$.

\begin{result}\label{res:PsAasmatrix}
The filter $P_h(A)$ with frequency response~$\widehat{h}$  is
\begin{align}\label{eqn:Ps(A)froms-2}
P_h(A){}&=\textrm{GFT}^{-1} \textrm{diag}\left[\widehat{h}\right]\textrm{GFT}.
\end{align}
\end{result}
\begin{proof} By~\eqref{eqn:graphfrequencyresponse-1a}, $\widehat{h}=P_h(\Lambda)\cdot 1$. Then, $P_h(\Lambda)=\textrm{diag}\left[\widehat{h}\right]$ and the result follows.
\end{proof}

We refer to $P_h(A)$ as the LSI filter associated with~$h$ or~$\widehat{h}$. Result~\ref{res:PsAasmatrix} gives $P_h(A)$ as a matrix filter. The next result gives it as a LSI or polynomial filter.
\begin{result}[LSI filter associated with~$\widehat{h}$]\label{res:PsA-1} Let the vector of coefficients of $P_h(A)$ be $p_h=\left[p_0\cdots p_{N-1}\right]^T$. Then, under assumption~\ref{assp:distincteigenA} of distinct eigenvalues, $p_h$ is the solution to
\begin{align}\label{eqn:graphsigimprespLSIfltr-3}
%    D^{\textrm{flat}}p_h{}&=h
%    \xrightarrow{F}
\mathcal{V}_\lambda p_h=\widehat{h}
\end{align}
where $\mathcal{V}_\lambda$ is the Vandermonde matrix of eigenvalues of~$A$:
\begin{align}
\label{eqn:matriximpv-5d}
\mathcal{V}_{\lambda}{}&=\left[\lambda^0\lambda^1\cdots\lambda^{N-1}\right]= \left[\begin{array}{cccc}
1&\lambda_0&\cdots&\lambda_0^{N-1}\\
\vdots&&&\vdots\\
1&\lambda_{N-1}&\cdots&\lambda_{N-1}^{N-1}
\end{array}\right]
\end{align}
\end{result}
Equation~\eqref{eqn:matriximpv-5d} defines the vector~$\lambda$ of eigenvalues and its powers $\lambda^n$ as Hadamard products of itself
\vspace*{-.4cm}
\begin{align}
\label{eqn:matriximpv-5d-1}
\lambda{}&=\left[1\,\lambda_0\,\lambda_1\hdots\lambda_{N-1}\right]^T, \textrm{  and   } \lambda^n=\overbrace{\lambda\odot\hdots\odot \lambda}^{\textrm{$n$ times}}.
\end{align}
\begin{proof}
The proof follows because $P_h(\Lambda)\cdot 1=\mathcal{V}_{\lambda}\cdot p$.
\end{proof}
For DSP, the Vandermonde is (apart a normalizing factor) the DFT matrix, and~\eqref{eqn:graphsigimprespLSIfltr-3} shows that in DSP $p_h$ is the inverse $\textrm{DFT}$ of~$\widehat{h}$, i.e., the impulse response of $P(A)$.
	\section{$\textrm{GSP}_{\textrm{sp}}$: Spectral Shift and Spectral GSP}
	\label{sec:spectralshift}
To show for GSP sampling the vertex/ spectral dualism in DSP sampling and to show the explicit  relationship between graph sampling in the vertex and in the graph spectral domains, we introduce graph convolution or graph filtering in the \textit{spectral} domain. In particular, we define linear shift invariant graph filters in the graph spectral domain as polynomials $P(M)$ of a spectral shift~$M$. The operator~$M$  shifts the graph spectrum~$\widehat{s}$ of a signal~$s$. This section establishes such an~$M$.

We start by recalling the DSP properties of shifting a signal in the time and frequency domains \cite{oppenheimwillsky-1983}:%,oppenheimschaffer-1989}:
	\begin{align}
\label{eqn:shiftprop-vertex}
	s_{n-1} {}&\xrightarrow{\mathcal{F}} e^{-j\frac{2\pi}{N}m}\widehat{s}_{m}\\
    \label{eqn:shiftprop}
	e^{j\frac{2\pi}{N}n}s_n {}&\xrightarrow{\mathcal{F}} \widehat{s}_{m-1}.
	\end{align}
 These equations show that shifting in the time domain multiplies the Fourier coefficient $\widehat{s}_m$ by the eigenvalue $\lambda_m=e^{-j\frac{2\pi}{N}m}$ of the shift~$A$. Likewise, shifting in the frequency domain multiplies the signal sample $s_k$ by $\lambda_k^*=e^{j\frac{2\pi}{N}k}$, the complex conjugate of the eigenvalue $\lambda_k$ of~$A$. Collecting the time samples $s_n$ in the vector signal~$s$ and the Fourier coefficients $\widehat{s}_m$ in vector~$\widehat{s}$, equations~\eqref{eqn:shiftprop-vertex} and~\eqref{eqn:shiftprop} lead to
	\begin{align}
\label{eqn:shiftprop-vertex-22}
	A\cdot s {}&\xrightarrow{\mathcal{F}} \Lambda\cdot\widehat{s}\\
    \label{eqn:shiftprop-2}
	\Lambda^*\cdot s {}&\xrightarrow{\mathcal{F}} \left[\begin{array}{c}
    \widehat{s}_{N-1}\\
    \widehat{s}_{0}\\
%    \widehat{s}_{1}\\
    \vdots\\
    \widehat{s}_{(N-1)-1}\\
    \end{array}\right]=A\cdot \widehat{s}.
	\end{align}
Equation~\eqref{eqn:shiftprop-vertex-22} shows that shifting signal~$s$ in the vertex (time) domain multiplies $\widehat{s}$ by the diagonal matrix~$\Lambda$ of the eigenvalues of~$A$. Similarly, equation~\eqref{eqn:shiftprop-2} shows that shifting vector~$\widehat{s}$ in the frequency domain multiplies~$s$ by ~$\Lambda^*$, the diagonal matrix of the conjugate eigenvalues of~$A$.

The DFT of the left side of~\eqref{eqn:shiftprop-2} is its right-hand side. Inserting $\textrm{DFT}^H \cdot \textrm{DFT}$ as below, get
\begin{align}\label{eqn:shiftprop-3}
\hspace{-.3cm}\underbrace{\textrm{DFT}\cdot\Lambda^*\cdot\textrm{DFT}^H}_{M} \cdot \underbrace{\textrm{DFT}\cdot{s}}_{\widehat{s}}=\left[\begin{array}{c}
    \widehat{s}_{N-1}\\
    \widehat{s}_{0}\\
%    \widehat{s}_{1}\\
    \vdots\\
    \widehat{s}_{(N-1)-1}\\
    \end{array}\right]=
A\cdot\widehat{s}.
	\end{align}
The middle vector in~\eqref{eqn:shiftprop-3} is of course the shifted (by one) $\widehat{s}$ and the right-hand side equation in~\eqref{eqn:shiftprop-3} is the same as the right-hand side equation in~\eqref{eqn:shiftprop-2}. The `surprise' here is that the left-hand side of~\eqref{eqn:shiftprop-3} indicates that the shifted (by one) $\widehat{s}$ is also obtained by multiplying~$\widehat{s}$ by a new `spectral shift' $M$. We readily recognize that in this case $M=A^*$ and since~$A$ is real the DSP spectral shift is $M=A$.

Because it will be important in the sequel, we write together the dual pairs, equation~\eqref{eqn:shiftprop-vertex-22} that shifts in time, and the equation that shifts in frequency, resulting from combining equation~\eqref{eqn:shiftprop-2} and the left-hand side of~\eqref{eqn:shiftprop-3}:
	\begin{align}
\label{shifteqn-3}
A\cdot s {}&\xrightarrow{\mathcal{F}} \Lambda\cdot\widehat{s}\\
\label{shifteqn-2}
\Lambda^* s&\xrightarrow{\mathcal{F}} M\cdot\widehat{s}.
	\end{align}
%
%

%	Section~\ref{sec:primerongsp} builds GSP \cite{Sandryhaila:13,Sandryhaila:14,Sandryhaila:14big}, see also\cite{ortegafrossardkovacevicmouravandergheynst-2018}, from the shift~$A$ that shifts graph signals in the vertex domain. In many applications, we shift the graph signal in the graph spectral domain. We introduce in this section the \textit{spectral shift}~$M$ \cite{shimoura-asilomar2019,shi2019graph} and develop a dual GSP, $\textrm{GSP}_{\textrm{sp}}$.
	%
%	\vspace*{-.3cm}	
\subsection{Spectral shift~$M$}\label{subsec:spectralshift}
We define the graph spectral shift~$M$ in GSP so that~\eqref{shifteqn-3} and~\eqref{shifteqn-2} are preserved and remain invariant in GSP.
\begin{definition}[GSP: Spectral shift~$M$ \cite{shimoura-asilomar2019,shi2019graph}]\label{def:gspspectralshiftM}
Let the vertex graph shift~$A$ be diagonalized in~\eqref{Adecompgsp}, with $\Lambda$ the diagonal matrix of eigenvalues, and~$s$, $\widehat{s}$, and $y=\Lambda^* \widehat{s}$ be given. Then, the graph spectral shift~$M$ is the operator defined by
	\begin{align}
\label{eqn:shifteqnGSP-2}
\Lambda^* s&\xrightarrow{\mathcal{F}} M\cdot\widehat{s}.
	\end{align}
\end{definition}
By this definition, the DSP duality~\eqref{shifteqn-3} and~\eqref{shifteqn-2} holds for GSP shifting in the vertex and spectral graph domains.

The next result gives an explicit expression for~$M$.
\begin{result}[GSP: Spectral shift~$M$ \cite{shimoura-asilomar2019,shi2019graph}]\label{res:gspspectralshiftM}
The shift~$M$ is
	\begin{align} \label{eqn:Mgeqn}
	M {}&= \textrm{GFT} \cdot \Lambda^* \cdot \textrm{GFT}^{-1}.
	\end{align}
\end{result}
\begin{proof}
The proof mimics  the steps going from~\eqref{eqn:shiftprop-2} to~\eqref{eqn:shiftprop-3}.

\textbf{If}: Multiply on the left by $\textrm{GFT}^{-1}$ the right-hand side of~\eqref{eqn:shifteqnGSP-2} and insert $\textrm{GFT}\cdot\textrm{GFT}^{-1}$ between $M$ and~$\widehat{s}$
	\begin{align}
\label{eqn:shifteqnGSP-3}
\textrm{GFT}^{-1} \cdot M\cdot \textrm{GFT}\cdot\textrm{GFT}^{-1}\cdot\widehat{s}.
	\end{align}
Now replacing~$M$ in~\eqref{eqn:shifteqnGSP-3} by its expression in~\eqref{eqn:Mgeqn}, canceling terms, and recognizing that $\textrm{GFT}^{-1}\cdot\widehat{s}=s$, we get the left-hand side of~\eqref{eqn:shifteqnGSP-2} as desired.

\textbf{Only if}: Start from definition~\ref{def:gspspectralshiftM}. Multiply the left-hand side of~\eqref{eqn:shifteqnGSP-2} by $\textrm{GFT}$ and insert between $\Lambda^*$ and~$s$ the product $\textrm{GFT}^{-1}\!\!\cdot\textrm{GFT}$. Get
\vspace*{-.25cm}
\begin{align}\label{eqn:onlyifpart-1}
\underbrace{\textrm{GFT}\cdot\Lambda^*\cdot\textrm{GFT}^{-1}}_{Q}\cdot \underbrace{\textrm{GFT}\cdot s}_{\widehat{s}}=M\cdot \widehat{s}\\
\label{eqn:onlyifpart-2}
Q\cdot\widehat{s}=M\cdot \widehat{s}.
\end{align}
Since~\eqref{eqn:onlyifpart-1} holds for every~$s$, and so for every~$\widehat{s}$, conclude from~\eqref{eqn:onlyifpart-2} $M=Q=\textrm{GFT}\cdot\Lambda^*\cdot\textrm{GFT}^{-1}$, proving the result.
\end{proof}
%\begin{result}[$M$ as spectral shift]\label{res:Mactsasspectralshift}
%For time signals, the spectral shift~$M$ acts as a shift in the frequency domain.
%\end{result}
%%
%\begin{proof}
%
%
	%Applying~$M$ to $\widehat{s}$ $m_0$ times, \eqref{shifteqn-2} leads to~\eqref{eqn:shiftprop-2}.

%

Definition~\ref{def:gspspectralshiftM} generalizes~$M$ in DSP as presented in~\eqref{eqn:shiftprop-3} since the $\textrm{GFT}$ is the $\textrm{DFT}$ in DSP and the $\textrm{DFT}$ is unitary.
\begin{remark}\label{rem:twodefspshift}
Reference~\cite{leus2017dual} defines a different spectral shift, requiring it to satisfy a number of properties like permutation invariance, seldom verified.  The reference uses $\Lambda$ rather than $\Lambda^*$ in~\eqref{eqn:Mgeqn}. For example, with time signals, our GSP definition~\ref{def:gspspectralshiftM} reduces to the DSP definition in~\eqref{eqn:shiftprop-3} and leads to $M=A$ as observed before. In contrast, the spectral shift~$M^\prime$ in \cite{leus2017dual} leads to a spectral shift that is $M^\prime=A^T$. In DSP and time signals, $A^T\neq A$, $A^T$ reverses the direction of the cycle graph and of time.
\end{remark}
While for DSP, $M=A$, in GSP~$M$ may not equal~$A$ \cite{shimoura-asilomar2019,shi2019graph}. The next result addresses when $M=A$ for GSP.
\begin{result}[GSP: $A=M$]\label{res:GSPshiftsAM}
Let~$A$ be real and normal. Then $A=M$ if $\textrm{GFT}^{\,T}=\textrm{GFT}$.
\end{result}
\begin{proof}
For~$A$ normal, real, and $\textrm{GFT}$ symmetric,
\begin{align}\label{eqn:Anormalreal}
A=\textrm{GFT}^*\cdot\Lambda\cdot\textrm{GFT}=A^*=\textrm{GFT}\cdot\Lambda^*\cdot\textrm{GFT}^*=M,
\end{align}
since $\textrm{GFT}^{-1}=\textrm{GFT}^H$. This proves the result.
\end{proof}
\vspace*{-.25cm}
Result~\ref{res:GSPshiftsAM} is sufficient for $A=M$. If in addition the eigenvalues of~$A$ are nonzero, then we get a necessary condition.
\begin{result}[GSP: $A=M$]\label{res:GSPshiftsAM-2}
Let~$A$ be real, normal, and have nonzero eigenvalues. Then $A=M$ only if $\textrm{GFT}^{\,T}=\textrm{GFT}$.
\end{result}
\begin{proof}
Now, $M=A=A^*$ only if
\begin{align}\label{eqn:Anormalreal-2}
\textrm{GFT}\cdot\Lambda^*\cdot\textrm{GFT}^H=\textrm{GFT}^H\cdot\Lambda\cdot\textrm{GFT}=\textrm{GFT}^T\cdot\Lambda^*\cdot\textrm{GFT}^*
\end{align}
Multiply the leftmost and rightmost sides of the equation by \!$\textrm{GFT}^H$\!\! on the left and by $\textrm{GFT}$ on the right:
\begin{align}\label{eqn:Anormalreal-3}
\Lambda^*=\textrm{GFT}^H\cdot\textrm{GFT}^T\cdot\Lambda^*\cdot\textrm{GFT}^*\cdot \textrm{GFT}.
\end{align}
For all the eigenvalues nonzero, this holds only if
\begin{align}\label{eqn:Anormalreal-4a}
\textrm{GFT}^H\cdot\textrm{GFT}^T=I=\textrm{GFT}^*\cdot \textrm{GFT}.
\end{align}
Taking the transpose on the left, since $I^T=I$, get
\begin{align}\label{eqn:Anormalreal-4}
\textrm{GFT}\cdot\textrm{GFT}^*=I=\textrm{GFT}^*\cdot \textrm{GFT},
\end{align}
which is true only if $\textrm{GFT}^{-1}=\textrm{GFT}^*$. Since the inverse is unique, $\textrm{GFT}^H=\textrm{GFT}^*$, implying $\textrm{GFT}^T=\textrm{GFT}$.
\end{proof}

We define LSI filtering in the spectral domain.

\textbf{Product of signals and LSI spectral filtering}. LSI polynomial spectral filtering in the \textit{spectral} domain is matrix-vector multiplication of a polynomial filter $P(M)$
 \begin{align}\label{eqn:convspectral-11a}
	P(M) {}&=p_0I+p_1M+\cdots+p_{N-1}M^{N-1},
	\end{align}
 with vector~$\widehat{s}$
 \begin{align}\label{eqn:convtime-2nd}
	\widehat{t}{}&=P(M) \cdot \widehat{s}=\left[p_0I+p_1M+\cdots+p_{N-1}M^{N-1}\right]\cdot \widehat{s}.
	\end{align}

\textbf{Vertex domain product--spectral convolution}. Consider the duality between product and convolution.
\begin{result}[Vertex domain product--spectral convolution]\label{res:pointwiseprod-1}
\vspace*{-.2cm}
\begin{align}\label{eqn:P(M)widehats-1}
p\left(\lambda^*\right)\odot s=P\left(\Lambda^*\right)\cdot s&\xrightarrow{\mathcal{F}}P(M)\cdot \widehat{s}.
\end{align}
\end{result}
The proof follows from the eigendecomposition of $P(M)$.

\textbf{Graph `vertex' response}. In analogy to the frequency response of a LSI filter $P(A)$ given in~\eqref{eqn:graphfrequencyresponse-1a}, we let the ``vertex response'' of the filter $P(M)$ to be $p\left(\lambda^*\right)$ defined in~\eqref{eqn:P(M)widehats-1}.

\textbf{LSI filter $P_s(M)$ and its vertex response $s$}. 
Similarly to result~\ref{res:PsAasmatrix} and result~\ref{res:PsA-1}, we determine LSI filter $P_s(M)$ from ``vertex response'' $s$. Let the coefficients of $P_s(M)$ be $p_s=\left[p_0\hdots p_{N-1}\right]$.
\begin{result}[Matrix $P_s(M)$]\label{res:Ps(M)directfroms} The LSI filter $P_s(M)$ with vertex response~$s$ is
\begin{align}
    \label{eqn:Ps(M)froms-2}
P_s(M){}&=\textrm{GFT}\,\textrm{diag}\left[s\right]\,\textrm{GFT}^{-1}.
\end{align}
\end{result}
\begin{proof}The result follows from realizing that $P_s\left(\Lambda^*\right)=\textrm{GFT}^{-1} P_s(M)\textrm{GFT}=\textrm{diag}\left[s\right]$.
\end{proof}
\begin{result}[LSI spectral filter with vertex response~$s$] \label{res:LSIspectralfilterandvertexresponse}
Under assumption~\ref{assp:distincteigenA}, the vector~$p_s$ of coefficients of the LSI spectral filter $P_s(M)$ with vertex response~$s$ is
\begin{align}\label{eqn:Ps(M)froms}
\mathcal{V}^*_\lambda\cdot p_s{}&=s.
\end{align}
where the Vandermonde matrix $\mathcal{V}_\lambda$ is given in~\eqref{eqn:matriximpv-5d}.
\end{result}

%Equation~\eqref{eqn:Ps(M)froms-2} follows from~\eqref{eqn:graphsigimprespLSIfltr-4b} by observing

%\begin{align}
%   \label{eqn:graphsigimprespLSIfltr-4b}
%\xrightarrow{\mathcal{F}^{-1}}\frac{1}{\sqrt{N}}\mathcal{V}^*_\lambda p_s{}&=s
%\end{align}

\textbf{Spectral convolution}. To interpret steps in sampling in the spectral domain, we define convolution of two spectral signals. Let $\oast$ represent (circular) convolution. The next result tells us how to compute convolution.
\begin{result}[Spectral convolution of two signals]
\label{res:convolutionshatandthat}
Consider spectral signals~$\widehat{s}$ and~$\widehat{t}$ and their corresponding LSI filters $P_s(M)$ and $P_t(M)$, where~$M$ is the spectral shift.  Then
\begin{align}\label{eqn:convolutionPsMtimest}
    \widehat{u}{}&=\widehat{s}\oast \widehat{t}=P_s(M)\cdot \widehat{t}=P_t(M)\cdot \widehat{s}\\
    \label{eqn:convolutionPsMPtM}
    \widehat{u}{}&=\widehat{s}\oast \widehat{t}=P_s(M) P_t(M)\cdot \widehat{\delta}_{\textrm{sp},0}^{\textrm{flat}}\\
    %\vspace*{-.1cm}
    \label{eqn:convolutionthroughspfilter-2}
   \widehat{u}{}&=\widehat{s}\oast \widehat{t} \xleftarrow{\mathcal{F}} u=s\odot t
\end{align}
where we define the spectral domain impulse
%\begin{align}\label{eqn:implseflatGSPsp}
 $\widehat{\delta}_{\scriptsize\textrm{sp},0}^{\textrm{flat}}  \xleftrightarrow{\mathcal{F}} \delta_{\scriptsize\textrm{sp},0}^{\textrm{flat}}=\frac{1}{\sqrt{N}}1$, to be flat (constant) in the vertex domain.
%\end{align}
\end{result}
Equation~\eqref{eqn:convolutionPsMtimest} shows spectral convolution of~$\widehat{s}$ and~$\widehat{t}$ as filtering of~$\widehat{s}$ (or~$\widehat{t}$) with LSI filter $P_t(M)$ (or $P_s(M)$), while~\eqref{eqn:convolutionPsMPtM} shows spectral convolution of the two signals as the \textit{impulse response} of the LSI filter $P_s(M)P_t(M)$.  Equation~\eqref{eqn:convolutionthroughspfilter-2} shows spectral convolution in the spectral domain as pointwise multiplication in the vertex domain.
%If $A\neq M$, $A$ and~$M$ still share some properties.
%
%
%%%%%%%%%%%%%%%%%%%%%%%%%%%%%%%%%%%%%%%%%%%%%%%%%%%%%%%%%%%%%%%%%%%%%%%%%%%%%%%%%%%%%%%%%%%%%%%%%%%%%%%%%%%%%%%%%%%%%%%%%%%%%%%%%%%%%%%%%%%%%%%%%%%%%%%%%%%%%
%\textit{Sparsity}. Under result~\ref{res:GSPshiftsAM}, $A$ sparse will mean~$M$ is sparse. But in general for~$A$ sparse $M$ will not be sparse, even though many of  its entries will be small in magnitude. Similarly, if~$M$ is sparse, then~$A$ will have many entries small in magnitude.
%
\vspace{-.2cm}
\subsection{Vertex shift~$A$ and graph spectral shift~$M$: Equivariance}\label{subsec:equivariance}
For graph $G=(V,E)$, adjacency~$A$ is defined up to a relabelling of the vertices by permutation~$\Pi_1$. Denote a quantity with respect to a new relabeling by $(\cdot)^\prime$. Then
\begin{align}\label{eqn:relabeling-1}
A^\prime{}&=\Pi_1\cdot  A \cdot \Pi_1^T\\
\label{eqn:relabeling-1s}
s^\prime{}&=\Pi_1 \cdot s.
\end{align}
To see how~$\Pi_1$ affects the graph spectral shift~$M$, we first consider how~$\Pi_1$ impacts the $\textrm{GFT}$ and $\Lambda$.

Eigendecomposing $A^\prime$, from~\eqref{Adecompgsp}, get two forms
 \begin{align}\label{eqn:relabeling-5}
A^\prime{}&=\Pi_1\cdot\textrm{GFT}^{-1}\cdot\Pi_2^H\cdot\Pi_2\cdot\Lambda\cdot\Pi^H_2\cdot\Pi_2\cdot\textrm{GFT} \cdot \Pi_1^T\\
\label{eqn:relabeling-6}
{}&=\Pi_1\cdot\textrm{GFT}^{-1}\cdot\Lambda\cdot\textrm{GFT} \cdot \Pi_1^T.
\end{align}
where $\Pi_2$ is unitary. These lead to two alternatives
 \begin{align}\label{eqn:relabeling-3}
\textrm{GFT}^\prime{}&=\Pi_2\cdot\textrm{GFT} \cdot \Pi^T_1\\
\label{eqn:relabeling-4}
\textrm{GFT}^{\prime\prime}{}&=\textrm{GFT} \cdot \Pi^T_1,
\end{align}
with~$\Lambda$ is either one of the two:
 \begin{align}\label{eqn:relabeling-9}
\Lambda^\prime{}&=\Pi_2\cdot\Lambda \cdot \Pi^H_2\\
\label{eqn:relabeling-10}
\Lambda^{\prime\prime}{}&=\Lambda.
\end{align}
In \eqref{eqn:relabeling-9}, $\Pi_2$ must be a permutation matrix to keep $\Lambda^\prime$ diagonal with the same eigenvalues as~$\Lambda$. For each of the two definitions of the GFT, the graph Fourier transform $\widehat{s}$ of~$s$ after relabeling with~$\Pi_1$
 \begin{align}\label{eqn:relabeling-7}
\widehat{s}^\prime{}&=\textrm{GFT}^{\prime}\cdot s^\prime=\Pi_2\cdot\textrm{GFT} \cdot \Pi^T_1\cdot\Pi_1 \cdot s=\Pi_2 \cdot\widehat{s}\\
\label{eqn:relabeling-8}
\widehat{s}^{\prime\prime}{}&=\textrm{GFT}^{\prime\prime}\cdot s^\prime=\textrm{GFT} \cdot \Pi^T_1\cdot\Pi_1 \cdot s=\widehat{s}.
\end{align}
The first definition permutes the graph Fourier transform $\widehat{s}$ by~$\Pi_2$. The second leaves~$\widehat{s}$ invariant. The question is which $\textrm{GFT}$ should be adopted: \eqref{eqn:relabeling-3} or~\eqref{eqn:relabeling-4}.

To resolve this, we look at a simple DSP example.  Consider, for example, $N=3$,  $s=\left[s_0\,\,s_1\,\,s_2\right]^T$, and a circular shift of the nodes to get $s^\prime=\left[s_2\,\,s_0\,\,s_1\right]^T$. If we use $\textrm{GFT}^{\prime\prime}$ and $\Lambda^{\prime\prime}=\Lambda$ from~\eqref{eqn:relabeling-4} and~\eqref{eqn:relabeling-10},
\begin{align}
\label{eqn:timereshuffle-1}
\hspace{-5mm}\Lambda^*\!\cdot\! \Pi_1\!\cdot\! s{}&\!=\!\left[\!\begin{array}{ccc}
1&&\\
&e^{-j\frac{2\pi}{3}}&\\
&&e^{-j\frac{2\pi}{3}2}
\end{array}\!\right]\!\cdot\! \left[\!\begin{array}{c}
s_2\\
s_0\\
s_1
\end{array}\!\right]\!=\!\left[\!\begin{array}{c}
s_2\\
e^{-j\frac{2\pi}{3}}s_0\\
e^{-j\frac{2\pi}{3}2}s_1
\end{array}\!\right]\!.
\end{align}
 In~\eqref{eqn:timereshuffle-1}, the time samples are multiplied by the wrong phase shift, for example, time sample $s_2$ is multiplied by~$1$ instead of $e^{-j\frac{2\pi}{3}2}$. We now consider computing~\eqref{eqn:shiftprop}, using $s^\prime$ from~\eqref{eqn:relabeling-1s}, $\textrm{GFT}^\prime$ from~\eqref{eqn:relabeling-3} and $\Lambda^\prime$ from~\eqref{eqn:relabeling-9}. Get
\begin{align}
\label{eqn:timereshuffle-2}
\hspace{-5mm}\Pi_2\!\cdot\!\Lambda^*\!\cdot\!\Pi^T_2\!\!\cdot\! \Pi_1\! \cdot\! s{}&\!=\!\!\left[\!\!\hspace{-.1cm}\begin{array}{ccc}
e^{-j\frac{2\pi}{3}2}&&\\
&1&\\
&&e^{-j\frac{2\pi}{3}}
\end{array}\hspace{-.1cm}\!\!\right]\!\cdot\! \left[\!\!\begin{array}{c}
s_2\\
s_0\\
s_1
\end{array}\!\!\right]\!\!=\!\!\left[\!\!\hspace{-.1cm}\begin{array}{c}
e^{-j\frac{2\pi}{3}2}s_2\\
s_0\\
e^{-j\frac{2\pi}{3}}s_1
\end{array}\hspace{-.1cm}\!\!\right]^T\!\!\!\!\!.\end{align}
For the right hand side in~\eqref{eqn:timereshuffle-2} to be the correct shifts as shown, $\Pi_2^T\Pi_1 $ needs to cancel and $\Pi_2=\Pi_1$. We conclude that, after relabeling the vertices of the graph by~$\Pi_1$, the $\textrm{GFT}$ should be given by~\eqref{eqn:relabeling-3} and not by~\eqref{eqn:relabeling-4}, with $\Pi_1 = \Pi_2$. The eigenvalue matrix $\Lambda$ should also then be permuted as in~\eqref{eqn:relabeling-9}, with $\Pi_1 = \Pi_2$. Also, when $\Pi_1 = \Pi_2 = I$, we obtain $A^\prime = A$ and the original eigendecomposition of $A$. 
%This is intuitively satisfying since it preserves duality to shifting in both the vertex and graph spectral domains.

We can now determine how relabeling nodes impacts~$M$.
\begin{result}[Equivariance to permutation]\label{res:permuteM}
When nodes of~$G$ are permuted by~$\Pi$, shifts~$A$ and~$M$ are conjugated %\vspace{-.3cm}
\begin{align}\label{eqn:AMequivariance}
A^\prime=\Pi\cdot A\cdot \Pi^T\textrm{    and    } M^\prime= \Pi\cdot M \cdot \Pi^T.
\end{align}
\end{result}
\vspace{-.1cm}
\begin{proof}
The equivariance of~$A$ to permutation was already proven in \cite{Sandryhaila:13}. We now consider the equivariance of~$M$.

From definition~\ref{def:gspspectralshiftM}, the action of~$M$ on~$\widehat{s}$ is the vector $y=\Lambda^*\dot s$ in the vertex domain. This spectral shifting property multiplies the vertex domain component $s_n$ of~$s$ by the conjugate of the graph frequency $\lambda_n$. If we reshuffle the labeling of the nodes by $\Pi$, then $\Lambda^*$ is conjugated by~$\Pi$, i.e., $\Lambda^*$ is given by~\eqref{eqn:relabeling-9}, in order to preserve the spectral shifting property. This forces the graph Fourier transform $\textrm{GFT}$ to also be conjugated by $\Pi$ as given by~\eqref{eqn:relabeling-3}. Similarly, we can conclude that $\textrm{GFT}^{-1}$ is conjugated by~$\Pi$. Putting these together leads to the equivariance of~$M$ to permutation~$\Pi$ as asserted by the result.
\end{proof}
\vspace{-.2cm}
Result~\ref{res:permuteM} is pleasing, it shows that~$A$ and~$M$ are impacted similarly: both are equivariant to~$\Pi$.
\begin{remark}[Scrambling vertex and spectral domains] \label{rmk:labelingvertices-eigenvalues}
In DSP, time and frequencies are usually implicitly ordered. This is so natural that DSP seldom explicitly discusses the indexing of the time samples or the indexing of the Fourier coefficients: the signal $\left(s_0,s_1,\cdots,s_{N-1}\right)$ is an ordered $N$-tuple, exactly like $\left(\widehat{s}_0,\widehat{s}_1,\cdots,\widehat{s}_{N-1}\right)$ is an ordered $N$-tuple. So, it may seem strange that in GSP one needs to share the ordering adopted for the vertices of the graph and/or for the graph frequencies. But actually this should not surprise us. Even in DSP, there are applications where it is useful to permute signal samples either in time or frequency. One such early technique for securing voice communication used scrambling \cite{jayant1982analog,SakuraiKogaMuratani-1984}. %,DelReFantacciMaffucci-1989,AcharyaReddyKumar-2009}.
 In simple terms, speech samples are scrambled to change their order. At the receiver, a descrambling block is required to reorder the speech samples. Other secure communications use scrambling in the frequency domain, or in other transform domains. This is to illustrate that although time signals and their spectra are naturally ordered by the time and frequency indices, reordering or permuting the samples in time or frequency have found applications in DSP. The important point is that scrambling requires then a descrambling block. In other words, the transmitter and receiver have a way to share their labeling scheme of the signal samples or of the spectral samples. Likewise, in GSP, different researchers working with the same graph need to share their node labeling to make sense of the graph signal. Likewise, if they share the graph spectrum, they must also share their labeling of graph frequencies to know the $\textrm{GFT}^{-1}$ that inverts the spectrum\footnote{Note that the graph and values defined on the graph do not depend on indexing. These entities exist independently of the indexing. It is only when researchers use matrices (e.g., the adjacency matrix) to represent the graph and vectors (e.g., the graph signal) to represent graph values that they need to choose an indexing.}.
\end{remark}
\subsection{$\textrm{GSP}_{\textrm{sp}}$: Dual Graph Signal Processing}
 The shifts~$A$ and~$M$ play twin or dual roles; just as GSP is built from~$A$, we build a dual $\textrm{GSP}_{\textrm{sp}}$ from~$M$.

 \textit{Data and spectral graphs}. As adjacency matrices, $A$ and~$M$ define graphs: shift~$A$ determines the (data) graph~$G$ whose node~$n$ indexes the data sample $s_n$, while the spectral shift~$M$ defines a new graph, the \textit{spectral} graph $G_{\textrm{sp}}=\left(V_{\textrm{sp}},E_{\textrm{sp}}\right)$, whose node~$m$ (the graph frequency $\lambda_m$, $m=0\cdots N-1$) indexes the graph Fourier coefficient $\widehat{s}_m$ of the data. As shown at the beginning of this section~\ref{sec:spectralshift}, in DSP, $G$ and~$G_{\textrm{sp}}$ are cycle graphs\textemdash the time samples are indexed by the time ticks (vertices of~$G$), the spectral coefficients are indexed by the frequencies (vertices of~$G_{\textrm{sp}}$).

 \textit{$\textrm{GFT}_{\textrm{sp}}$: Spectral GFT}. Since~$M$ is diagonalized in~\eqref{eqn:Mgeqn}, %
\begin{align}\label{eqn:gfts-1}
\textrm{GFT}_{\textrm{sp}}{}&=\textrm{GFT}^{-1}.
\end{align}
The spectral $\textrm{GFT}_{\textrm{sp}}$ of the spectral graph signal $\widehat{s}$ is
\begin{align}\label{eqn:gfts-2}
\widehat{\widehat{s}}{}&=\textrm{GFT}_{\textrm{sp}}\cdot\widehat{s}= \textrm{GFT}^{-1}\cdot\widehat{s}=s.
\end{align}
\textit{LSI (linear shift invariant) spectral filters}: Since~$M$ is diagonalizable,  LSI spectral filters $P_{\textrm{sp}}\left(M\right)$ are polynomials\footnote{In the sequel, we will usually ignore the subindexing sp.}, see equation~\eqref{eqn:convspectral-11a} and result~\ref{res:pointwiseprod-1}.
	
%%%%%%%%%%%%%%%%%%%%%%%%%%%%%%%%%%%%
%
%
\begin{example}[Star graph]\label{exp:stargraphM}
	\label{subsec:propertiesM}
%	\textbf{Star Graph:}
Consider~$A$ for the star graph
\begin{equation} \label{eqn:starA}
    A =
\begin{bmatrix}
0&1_{N-1}\\
1_{N-1}&0_{(N-1)(N-1)}
\end{bmatrix}
\end{equation}
%A star graph has 1s in the entire first column and first row of $A$ (except $A_{00}$) and no self loops.
%
Its eigenvalues are $\pm\sqrt{N-1}$ with multiplicity~1 and~0 with algebraic and geometric multiplicities~$N-2$.
%
%From the eigendecomposition of $A$ in~\eqref{eqn:starA}.
%
The $\textrm{GFT}$
\begin{align} \label{eqn:stargft}
    \textrm{GFT} {}&= \frac{1}{\sqrt{N-1}}
    {\scriptsize
\left[\hspace{-.2cm}
\begin{array}{cccccc}
 \frac{\sqrt{N-1}}{\sqrt{2}} & \hspace{-.2cm}-\frac{\sqrt{N-1}}{\sqrt{2}} & 0 & \hdots & 0\\
 -\frac{1}{\sqrt{2}} & \frac{1}{\sqrt{2}} & 1 & \hdots & 1 \\
 - \frac{1}{\sqrt{2}} & \frac{1}{\sqrt{2}} & e^{-j\frac{2\pi}{N-1}} & \hdots & e^{-j\frac{2\pi(N-2)}{N-1}}\\
%-\frac{1}{\sqrt{2}} & \frac{1}{\sqrt{2}} & e^{-j\frac{2\pi2}{N-1}}  &\hdots & e^{-j\frac{2\pi2(N-2)}{N-1}}\\
  \vdots & \vdots & \vdots& \ddots & \vdots \\
 -\frac{1}{\sqrt{2}} & \frac{1}{\sqrt{2}} & e^{-j\frac{2\pi(N-2)}{N-1}} &  \hdots & e^{-j\frac{2\pi(N-2)^2}{N-1}}\\
\end{array}
\hspace{-.2cm}\right]
}
\\
%\end{equation}
%The determinant of $\text{GFT}^{-1}$ is $2(N-1)^{\frac{3}{2}}$.
%By inverting $\text{GFT}^{-1}$, we obtain the GFT
%\begin{equation}
\label{eqn:stargfti}
%\textrm{GFT}^H{}& =
%{\scriptsize
%\frac{1}{\sqrt{N-1}}
%\left[
%\begin{array}{cccccc}
% \frac{\sqrt{N-1}}{\sqrt{2}} & \frac{1}{\sqrt{2}}  & \frac{1}{\sqrt{2}} & \hdots &  \frac{1}{\sqrt{2}}\\
% \frac{-\sqrt{N-1}}{\sqrt{2}} & \frac{1}{\sqrt{2}}  & \frac{1}{\sqrt{2}} & \hdots & \frac{1}{\sqrt{2}}\\
% 0 & 1 & e^{j\frac{2\pi}{N-1}} & \hdots & e^{j\frac{2\pi(N-2)}{N-1}}\\
%0 & 1 & e^{j\frac{2\pi(2)}{N-1}}  &\hdots & e^{j\frac{2\pi(2)(N-2)}{N-1}}\\
%  \vdots & \vdots & \vdots& \ddots & \vdots \\
%0 & 1 & e^{j\frac{2\pi(N-2)}{N-1}} &  \hdots & e^{j\frac{2\pi(N-2)^2}{N-1}}\\
%\end{array}
%\right]
%}
\textrm{GFT}_\textrm{sp}{}&=\textrm{GFT}^{-1}=\textrm{GFT}^H
\end{align}
This $\textrm{GFT}$ diagonalizes~$A$. To develop it, note that the characteristic polynomial of~$A$ is $\Delta(\lambda)=\lambda^{N-2}\left(\lambda-(N-1)\right)^2$ and that the eigenvectors of~$A$ are the same as $A^2$. Matrix~$A^2$ is block diagonal, with a scalar minor $N-1$ and a matrix minor $1_{N-1}\cdot 1_{N-1}^T$. The latter is circulant and diagonalized by $\textrm{DFT}_{N-1}$.
The spectral shift\footnote{Since the ordering of the nodes is specified, the $\textrm{GFT}$ and the spectral shift~$M$ are fixed.} for the star graph follows, shown for $N=5$ in  Figure~\ref{fig:star}:
\begin{align}\label{eqn:starM}
    M=\frac{1}{\sqrt{N-1}}\!
    \!\left[\!\!\!\begin{array}{lcl}
    \frac{N}{2}\left[\!\begin{array}{r}
    1\\
    -1
    \end{array}\!\right]\!\!
    \left[\!\!\begin{array}{cc}
    1&
    -1
    \end{array}\!\!\right]\!-\!I_2&&\!\!-\frac{1}{\sqrt{2}}\!\left[\!\begin{array}{c}
    1\\
    1
    \end{array}\!\right]\!\otimes\! 1_{N-2}^T\\
    &\\
    -\frac{1}{\sqrt{2}}\!\left[\!\begin{array}{cc}
    1&
    1
    \end{array}\!\right]\!\otimes\! 1_{N-2}&&\!\!-1_{N-2}\cdot 1_{N-2}^T
    \end{array}\!\right]
\end{align}
%
%
%
%\vspace{-.4cm}
\begin{figure}[hbt]
    \centering
    \includegraphics[width=6.0cm,keepaspectratio]{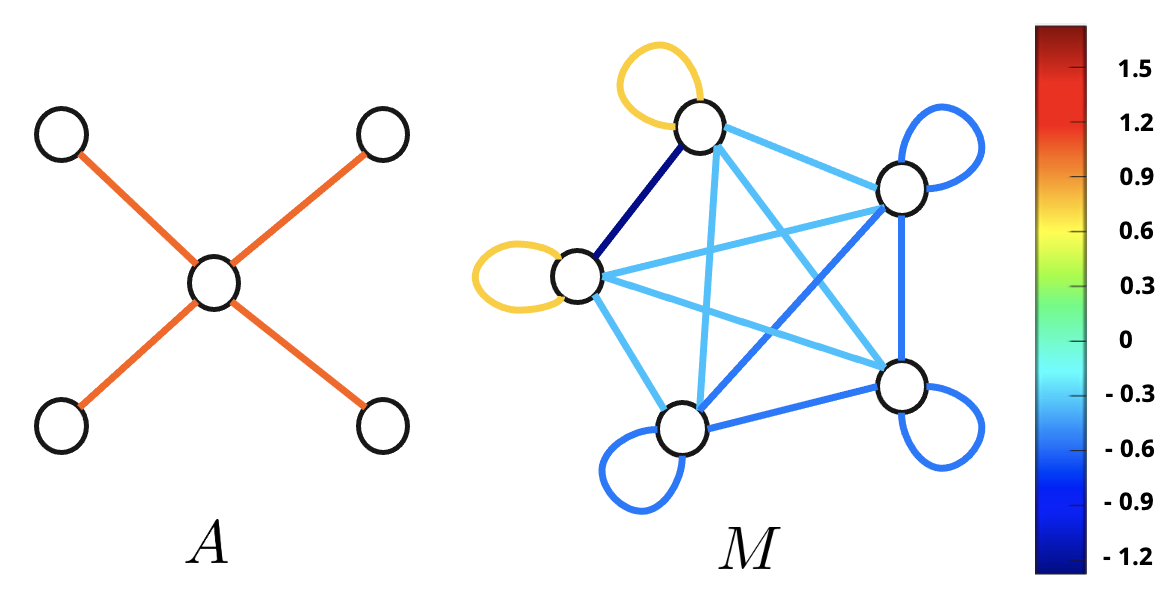}
    \caption{Star graph: Shifts~$A$ and~$M$ for $N=5$.}
    \label{fig:star}
\end{figure}
%Except for a few, the entries of~$M$ decay as $1/\sqrt{N-1}$.
\vspace{-.3cm}

%We briefly consider the $A$ and $M$ for other graphs, shown in Figures \ref{fig:wheel} and \ref{fig:ladder}.
%\begin{figure}
%    \centering
%    \includegraphics[width=8cm,keepaspectratio]{images/wheel.png}
%    \caption{The adjacency matrix $A$ and corresponding $M$ for a wheel graph}
%    \label{fig:wheel}
%\end{figure}
%\begin{figure}
%    \centering
%    \includegraphics[width=7cm,keepaspectratio]{images/ladder.png}
%    \caption{The adjacency matrix $A$ and corresponding $M$ for a ladder graph}
%    \label{fig:ladder}
%\end{figure}
\end{example}

\vspace{-.2cm}	
\section{GSP Sampling: Subsampling and Decimation}\label{sec:subsamplingdecimation}
In DSP, sampling of bandlimited signals~$s$ include: \begin{inparaenum}[1)]
	\item\label{inp:downsampling-2} \textit{Subsampling} that zeroes samples of~$s$ (e.g., every other sample) to get a subsampled signal $s_\delta$;
	\item\label{inp:decimation-2} \textit{decimating} or \textit{downsampling} that discards zeroed samples in $s_\delta$ to get the decimated signal $s_d$; and
	\item\label{inp:reconstruction} \textit{reconstruction} that
	\begin{inparaenum}[3.i)]
	\item \textit{upsamples} $s_d$ by reinserting the zeros discarded to get back the upsampled signal $s_\delta$, and
	\item \textit{interpolates} by ideal \textit{lowpass filtering} $s_\delta$ to get\footnote{\label{ftn:aliasing} If sampling below the Nyquist rate, $s_r$ in 3.ii) is an aliased version of~$s$.} $s_r\!\!=\!\!s$.
	\end{inparaenum}
	\end{inparaenum}
 All these steps have interpretations in both time and frequency.

  Likewise, we consider the equivalent sequence of steps in GSP sampling\textemdash \textit{subsampling}, \textit{decimation}, and \textit{reconstruction}\textemdash and develop for each step dual interpretations in the vertex and graph spectral domains. To achieve these dual interpretations, we extend current GSP theory with new concepts and constructs. Further, we show explicitly which choices among alternatives have to be made in GSP sampling to obtain DSP sampling when the graph is the cyclic graph. These choices are implicit and taken for granted in DSP. Our work provides further insights and interpretations into DSP.

In this and the next section, $A$ is the adjacency matrix of a~$N$ node arbitrary graph.  Let $\left\|\widehat{s}\right\|_0$ be the $\ell_0$ pseudo-norm,  i.e., the number of nonzero entries of $\widehat{s}$. Let $s$ be bandlimited with bandwidth $K$, i.e., $\left\|\widehat{s}\right\|_0 \leq K$, $K \leq N$. For ease of notation, we assume the last $N-K$ entries of $\widehat{s}$ are zero, i.e., $\widehat{s} = [\widehat{s}_K^T\,\, \widehat{s}_{N-K}^T]^T$, where $\widehat{s}_{N-K} = 0$,\footnote{\label{ftn:shapeofbdlm-s} In actuality, the zero entries can occur anywhere in $\widehat{s}$. In this case, we are assuming that we reorder $\widehat{s}$ such that its last $N-K$ entries are 0.} and that~$K$ divides~$N$, $K|N$.

This section considers first the sampling set in subsection~\ref{subsec:samplingset}, subsampling in subsection~\ref{subsec:lsisampling}, and decimation in subsection~\ref{subsec:decimation}. In all these subsections, we provide vertex and spectral domain GSP interpretations that parallel DSP.

\subsection{Sampling Set~$S$}
\label{subsec:samplingset}
We refer to section~\ref{sec:introduction} that reviews the significant work in defining the sampling set~$S$. In our context, with $K|N$ and finite graphs, the sampling set~$S$ \cite{pesenson2008sampling,pesenson2001sampling,gadde2015probabilistic} is the minimum set of vertices indexing the signal samples that enables perfect reconstruction of bandlimited signals from the corresponding decimated signal~$s_d$.

\textbf{Choice of sampling set~$S$}:
We assume that the graph signal~$s$ is lowpass and bandlimited to~$K$ and $K|N$. To fix notation, we briefly describe one method to determine the sampling set~$S$, or its characteristic graph signal $\delta^{\scriptsize\textrm{(spl)}}$, a vector of zeros and ones. When~$s$ is bandlimited, knowing~$s$ at the vertices in~$S$ allows for perfect reconstruction of~$s$. This is the decimated or downsampled version $s_d$ of~$s$.
\begin{result}[Sampling set]\label{res:samplingset-1a}
With the notation and assumptions above, let graph $G=(V,E)$, $|V|=N$ whose nodes index bandlimited graph signals~$s$ with bandwidth~$K$, $K|N$. Then, there is a sampling set~$S$ with cardinality~$K$ and indicator signal $\delta^{\scriptsize\textrm{(spl)}}$ such that~$s$ is perfectly reconstructed from its samples indexed by vertices in~$S$.
\end{result}
\begin{proof}
The proof can be found in \cite{pesenson2008sampling,pesenson2010sampling}. In our paper, we consider one method of finding a sampling set and comment on its nonuniqueness.

We start with the Fourier relation between the signal~$s$ and its graph Fourier transform~$\widehat{s}$ and block partition rowwise the GFT matrix as indicated below:
	\begin{align}\label{eqn:GFTs-hats-1}
	\textrm{GFT}\, s{}&=\widehat{s}
	\:\:\Longrightarrow \hspace{-.8cm}\Aboxed{\left[\begin{array}{c}
	\textrm{GFT}_K\\
	\textrm{GFT}_{N-K}
	\end{array}\right]s=\left[\begin{array}{c}
	\widehat{s}_K\\
	\widehat{s}_{N-K}
	\end{array}\right]}
	\end{align}
with the top~$K$ rows of the $\textrm{GFT}$ in $\textrm{GFT}_K\!:\! K\times N$ and the bottom $N-K$ rows in $\textrm{GFT}_{N-K}\!:\! (N-K)\times N$. Given that $\textrm{GFT}$ is full rank, $\textrm{GFT}_K$ and $\textrm{GFT}_{N-K}$ are full rank.

%Equation~\eqref{eqn:GFTs-hats-1} is a linear system of equations with~$N$ unknowns, the entries of~$s$. Given $\widehat{s}$, the system~\eqref{eqn:GFTs-hats-1} determines in a unique way the signal~$s$ through $\textrm{GFT}^{-1}$. Given the bandlimitedness of $\widehat{s}$, $\widehat{s}_{N-K} = 0$, the top equation solves for~$s$ using the pseudo inverse of $\textrm{GFT}_K$. But $\widehat{s}_{N-K} = 0$ reduces the number of degrees of freedom of~$s$ from~$N$ to~$K$. Hence, what is of interest in sampling is to find a set of components of~$s$, the ``free'' variables, from which to determine the original~$s$.

Taking $\widehat{s}_{N-K} = 0$ in~\eqref{eqn:GFTs-hats-1}, we get
    \begin{equation}\label{eqn:freepivotvariablessystem-1}
	\text{GFT}_{N-K} s = \widehat{s}_{N-K} = 0.
	\end{equation}
%	Since rank of $\text{GFT}_{N-K}$ is $N-K$, this system of $N-K$ linearly independent equations in~$N$ unknowns (components of~$s$) imposes $N-K$ constraints on~$s$, limiting~$s$ to a~$K$ linear subspace.
Solving~\eqref{eqn:freepivotvariablessystem-1} determines $N-K$ components of~$s$ (so called pivot variables) in terms of the other~$K$ components (so called free variables). There are many alternative possible sets of $N-K$ pivots and $K$~free variables, i.e., this split is not unique. There are also different ways to determine these sets. %To be concrete,
We illustrate with Gauss Elimination~(GE) as just one method.

GE determines $N-K$ linearly independent rows and columns of $\textrm{GFT}_{N-K}$, reducing it to row echelon form:
	\begin{align}\label{eqn:GFT(N-K)GE-1}
\hspace{-.55cm}
\textrm{GFT}_{N-K}s{}&\!=\!\widehat{s}_{N-K}\!\xrightarrow{\textrm{GE}}\! E\cdot\textrm{GFT}_{N-K}\Pi^T{}_{\!\!\!\!\textrm{col}}\cdot\Pi_{\textrm{col}}s\!=\! 0.
	\end{align}
In~\eqref{eqn:GFT(N-K)GE-1}, $E$ represents the row operations that reduce the $\textrm{GFT}_{N-K}$ to row echelon form. Matrix $E\!:\! (N-K)\times (N-K)$ is the product of elementary matrices and so it is full rank. Partition the row echelon form of $\textrm{GFT}_{N-K}$ as
\begin{align}\label{eqn:defineBij}
E\cdot\textrm{GFT}_{N-K}\Pi^T_{\textrm{col}}{}&=\left[B_{11}\,B_{12}\right],
\end{align}
    with  $B_{11}\!:\! (N-K)\times (N-K)$ upper triangular with ones on the diagonal, and $B_{12}\!:\!(N-K)\times K$. The matrix~$\Pi_{\textrm{col}}$ is a permutation representing possible column swapping.

Let
\begin{align}\label{eqn:pivotfreevariables-1}
\Pi_{\textrm{col}}s{}&=\left[\begin{array}{c}
s_{N-K}\\
s_K
\end{array}\right].
\end{align}
Replacing~\eqref{eqn:defineBij} in~\eqref{eqn:GFT(N-K)GE-1} and using the partitioning of~$\Pi_{\textrm{col}}s$ in~\eqref{eqn:pivotfreevariables-1}, equation~\eqref{eqn:GFT(N-K)GE-1} becomes:
\begin{align}\label{eqn:GFT(N-K)GE-2}
\left[B_{11}\,B_{12}\right]\left[\begin{array}{c}
s_{N-K}\\
s_K
\end{array}\right]{}&=0.
\end{align}
Since $B_{11}$ is invertible, \eqref{eqn:GFT(N-K)GE-2} leads to
\begin{align}\label{eqn:GFT(N-K)GE-2a}
s_{N-K}{}&=-B_{11}^{-1}\cdot B_{12}\cdot s_K.
\end{align}
 This determines $s_{N-K}$ from $s_K$. The vector $s_{N-K}$ collects the $N-K$ pivot entries and $s_K$ collects the~$K$ free variables. This shows that, given the free variables $s_K$, we recover:
\begin{equation}\label{eqn:vertexgraphsampling-1}
    \Pi_{\textrm{col}}s = \left[\begin{array}{c}
    -B_{11}^{-1}\cdot B_{12}\\
    I_K
    \end{array}\right] s_K.
\end{equation}
  With~$S$ the set of indices of the free variables $s_K$ ($|S| = K$) and $\delta^{\scriptsize\textrm{(spl)}}$ its indicator signal, the result follows.
  %Now, $|S|=K$ since if $|S|>K$ ($|S|<K$) there would be more (fewer) than~$N-K$ linearly independent equations in~\eqref{eqn:freepivotvariablessystem-1}, contradicting that $\textrm{rank}\left(\text{GFT}_{N-K}\right)=N-K$, proving the result.
\end{proof}
\begin{remark}[$S$ not unique]\label{rmk:freevariables}
Applying GE to~\eqref{eqn:freepivotvariablessystem-1}, we can permute rows and columns of $\textrm{GFT}_{N-K}$, leading to different choices of pivots and free variables. Hence, $S$ and $\delta^{\scriptsize\textrm{(spl)}}$ are not unique. Regardless of the choice for~$S$,  $|S|=\left\|\delta^{\scriptsize\textrm{(spl)}}\right\|_0=K$ equals the number of degrees of freedom in~\eqref{eqn:freepivotvariablessystem-1} and the bandwidth of~$s$.
\end{remark}
\begin{result}[$S$ and $\delta^{\scriptsize\textrm{(spl)}}$]\label{res:Sandsamplingdelta}
Under the set-up of result~\ref{res:samplingset-1a}, given the sampling set~$S$, the sampling signal $\delta^{\scriptsize\textrm{(spl)}}$ is unique and the signal samples indexed by~$S$ uniquely determine~$s$.
\end{result}
This result is of course tautologic since $\delta^{\scriptsize\textrm{(spl)}}$ is the characteristic signal of~$S$ and~\eqref{eqn:vertexgraphsampling-1} shows how to recover~$s$ from $s_K$. We make it explicit for easy future reference that the degrees of freedom are in choosing~$S$. Once chosen, the sampling signal is fixed and~$s$ is uniquely determined.
\subsection{GSP Subsampling by LSI Filtering}
\label{subsec:lsisampling}
This section shows that GSP and DSP subsampling have equivalent vertex and spectral domain dual interpretations.

Assume sampling set~$S$ and its sampling graph signal $\delta^{\scriptsize\textrm{(spl)}}$ have been chosen. In DSP, uniform ideal subsampling~$s$ in the vertex domain is multiplication of~$s$ by a train of pulses $\delta^{\scriptsize\textrm{(spl)}}$. In the spectral domain, it is convolution (or LSI filtering) by a periodic train of pulses. We now discuss subsampling in GSP.

Let $G=(V,E)$, with shift~$A$ and spectral shift~$M$. Let $s_\delta$ be the subsampled graph signal obtained from~$s$. Its $K$ nonzero entries are the entries indexed by vertices in~$S$.
\begin{result}[GSP subsampling as LSI filter]\label{res:gspsamplingasLSI-1} Under assumption~\ref{assp:distincteigenA}, GSP subsampling in the spectral domain is LSI filtering
	\begin{align} \label{eqn:gspvertexsampling-1}
	s_\delta{}&=\delta^{\scriptsize\textrm{(spl)}} \odot s  \xrightarrow{\mathcal{F}} P_{\delta^{\scriptsize\textrm{(spl)}}}(M) \cdot \widehat{s}=\widehat{s}_\delta
	\end{align}
\end{result}
\begin{proof} Subsampling is pointwise multiplication
\begin{align}\label{eqn:GSPsampling-a1}
s_\delta{}&=\delta^{\scriptsize\textrm{(spl)}} \odot s
\end{align}
in the vertex domain, which is the left-hand side of~\eqref{eqn:gspvertexsampling-1}.

To show it is LSI filtering in the spectral domain, we need to show that the $\textrm{GFT}$ of $s_\delta$ is obtained by polynomial filtering $\widehat{s}$. By result~\ref{res:convolutionshatandthat} and equation~\eqref{eqn:convolutionthroughspfilter-2}, pointwise multiplication of $\delta^{\scriptsize\textrm{(spl)}}$ and~$s$ in the vertex domain is convolution in the graph spectral domain. By~\eqref{eqn:convolutionPsMtimest}, the convolution is filtering $\widehat{s}$ with the LSI polynomial filter $P_{\delta^{\scriptsize\textrm{(spl)}}}(M)$. We only need to show that, given the sampling signal $\delta^{\scriptsize\textrm{(spl)}}$,  $P_{\delta^{\scriptsize\textrm{(spl)}}}(M)$ is well defined.

By result~\ref{res:Ps(M)directfroms} and equation~\eqref{eqn:Ps(M)froms-2},  $P_{\delta^{\scriptsize\textrm{(spl)}}}(M)$ is given by
	\begin{align} \label{eqn:gspLSIsamp-1}
	P_{\delta^{\scriptsize\textrm{(spl)}}}(M) &= \textrm{GFT}\, \textrm{diag}\left[\delta^{\scriptsize\textrm{(spl)}}\right] \textrm{GFT}^{-1},
%	\\
%\label{gspsamp-2}
%	{}&=  \textrm{GFT} \ \textrm{diag}\left[\delta^{\scriptsize\textrm{(spl)}}\right]\ \textrm{diag}\left[\delta^{\scriptsize\textrm{(spl)}}\right] \ \textrm{GFT}^{-1}
	\end{align}
with coefficients $p_{\delta^{\scriptsize\textrm{(spl)}}}$ given by equation~\eqref{eqn:Ps(M)froms}:
\begin{align}\label{eqn:vertexsampling-LSIfilter-1}
\mathcal{V}_\lambda^*\cdot p_{\delta^{\scriptsize\textrm{(spl)}}}{}&=\delta^{\scriptsize\textrm{(spl)}}.
\end{align}
By assumption~\ref{assp:distincteigenA} of distinct eigenvalues, the Vandermonde matrix $\mathcal{V}_\lambda$ is full rank  and~\eqref{eqn:vertexsampling-LSIfilter-1} has a unique solution:
\begin{equation}\label{eqn:vertexsampling-LSIfilter-1v2}
    p_{\delta^{\textrm{(spl)}}}=\mathcal{V}_\lambda^{*{-1}}\delta^{\textrm{(spl)}}
\end{equation}
Hence, $p_{\delta^{\scriptsize\textrm{(spl)}}}$ and $P_{\delta^{\scriptsize\textrm{(spl)}}}(M)$ are well defined, proving the result.
%
%We then have as desired that subsampling in the graph spectral domain is LSI filtering by $P_{\delta^{\scriptsize\textrm{(spl)}}}(M)$
%\begin{align} \label{eqn:gspvertexsampling}
%	s_\delta{}&=\delta^{\scriptsize\textrm{(spl)}} \odot s  \xrightarrow{\mathcal{F}} P_{\delta^{\scriptsize\textrm{(spl)}}}(M)\cdot \widehat{s}=\widehat{s}_\delta
%	\end{align}
%
%\vspace*{-.4cm}
\end{proof}
%
%
          %becomes the DSP convolution in the frequency domain with trains of impulses when the graph is the DSP cycle graph.

      In DSP, it is well known that the spectrum of the subsampled signal $s_\delta$ is the nonzero spectrum $\widehat{s}_K$ of~$s$ replicated $N-K$ times. In other words, in the spectral domain DSP subsampling is convolution with a train of equispaced spectral pulses. We now wish to show that the graph spectrum $\widehat{s}_\delta$ of the subsampled signal $s_\delta=\delta^{\scriptsize\textrm{(spl)}} \odot s$ is given by (possibly filtered) copies of the nonzero spectrum of~$s$, i.e., $\widehat{s}_K$. At first sight, this is not obvious. In fact, the spectral domain LSI filter  $P_{\delta^{\scriptsize\textrm{(spl)}}}(M)$ with coefficients $p_{\delta^{\scriptsize\textrm{(spl)}}}=\left[p_0\,\,p_1\,\,\cdots p_{N-1}\right]^T$ given by~\eqref{eqn:vertexsampling-LSIfilter-1} is
     \begin{align}\label{eqn:polyP(M)-a1}
     P_{\delta^{\scriptsize\textrm{(spl)}}}(M)=p_0I+p_1M+\cdots+p_{N-1}M^{N-1}.
     \end{align}
     Then,
     \begin{align}\label{eqn:polyP(M)-a2}
     \widehat{s}_\delta{}&=\left(p_0I+p_1M+\cdots+p_{N-1}M^{N-1}\right)\left[\begin{array}{c}
     \widehat{s}_K\\
     0_{N-K}
     \end{array}\right].
     \end{align}
      This shows $\widehat{s}_\delta$ is a superposition of replicas of $\widehat{s}_K$, which could overlap. We show this is not the case, if the bandlimited graph signal is sampled at graph Nyquist rate~$K$.

      \begin{result}[Replication: GSP spectrum of subsampled $s_\delta$] \label{res:GSPspectrumsubsampledsig}
      The spectrum $\widehat{s}_\delta$ of the subsampled $s_\delta$ corresponds to $\frac{N}{K}$ (filtered, possibly distorted) copies of the nonzero spectrum $\widehat{s}_K$ of $s$. Sampling at the ``graph'' Nyquist rate~$K$, aliasing does not occur.
      \end{result}
      We start with preliminary notation before the proof. Assume the sampling set~$S$ has been chosen with given sampling graph signal $\delta^{\scriptsize\textrm{(spl)}}$. Without loss of generality, to make the presentation easier, assume reordering the vertices of the graph by permutation~$\Pi$ so that
\begin{align}\label{eqn:samplingdelta}
\delta^{\scriptsize\textrm{(spl)}}=\left[1_K^T 0^T_{N-K}\right]^T.
\end{align}
Permutation~$\Pi$ conjugates~$A$, $\textrm{GFT}$, and $\textrm{GFT}^{-1}$. We will ignore~$\Pi$. Partition $\textrm{GFT}$, $\textrm{GFT}^{-1}$, and $P_{\delta^{\scriptsize\textrm{(spl)}}}(M)$:
    \begin{align}
        \label{eqn:partitionGFTcl-1}
	\textrm{GFT}\phantom{^{-1}} {}&= \begin{bmatrix}
	\textrm{GFT}_{K} &\hspace{+.1cm} \textrm{GFT}_{N-K}\hspace{.04cm}
	\end{bmatrix}\\
    \label{eqn:partitionGFTcl-0}
	{}&=\left[\!\!\begin{array}{ll}
	\textrm{GFT}_{KK} \!&\! \textrm{GFT}_{K(N-K)}\\
	\textrm{GFT}_{(N-K)K} \!&\! \textrm{GFT}_{(N-K)(N-K)}
    \end{array}\!\!\right]\\
    \label{eqn:partitionGFTcl-2}
	\textrm{GFT}^{-1} {}&= \begin{bmatrix}
	\textrm{GFT}^{-1}_{K} & \hspace{+.08cm}\textrm{GFT}^{-1}_{N-K}
	\end{bmatrix}\\
    \label{eqn:partitionGFTcl-3}
	{}&=\left[\!\!\begin{array}{ll}
	\textrm{GFT}^{-1}_{KK} \!&\! \textrm{GFT}^{-1}_{K(N-K)}\\
	\textrm{GFT}^{-1}_{(N-K)K} \!&\! \textrm{GFT}^{-1}_{(N-K)(N-K)}
    \end{array}\!\!\right]\\
    \label{eqn:partitionGFTcl-4}
	P_{\delta^{\scriptsize\textrm{(spl)}}}(M) {}&= \begin{bmatrix}
	P_{\delta^{\scriptsize\textrm{(spl)}}}(M)_{K} &\hspace{-.15cm} P_{\delta^{\scriptsize\textrm{(spl)}}}(M)_{N-K}\hspace{-.03cm}
	\end{bmatrix}.
    \end{align}
    	In~\eqref{eqn:partitionGFTcl-1}, \eqref{eqn:partitionGFTcl-2}, and~\eqref{eqn:partitionGFTcl-4}, the partitions are columnwise, not rowwise as in~\eqref{eqn:GFTs-hats-1} and the left blocks are $N\times K$ and the right blocks $N\times (N-K)$. In~\eqref{eqn:partitionGFTcl-0} and \eqref{eqn:partitionGFTcl-3}, the subindices give the dimensions of each subblock of $\textrm{GFT}$ and $\textrm{GFT}^{-1}$. E.g., $\textrm{GFT}_{KK}$ is the top left $K \times K$ subblock of the GFT.
      \begin{proof}
      Recall the filter $P(M)_{\delta^{\scriptsize\textrm{(spl)}}}$ given by~\eqref{eqn:gspLSIsamp-1}.
%
%
%	By result~\ref{res:gspsamplingasLSI-1} and~\eqref{eqn:gspvertexsampling-1}, $\delta^{\scriptsize\textrm{(spl)}}\odot s\xrightarrow{\mathcal{F}} P(M)\cdot\widehat{s}$, with
Using~\eqref{eqn:partitionGFTcl-1} and~\eqref{eqn:partitionGFTcl-2} in~\eqref{eqn:gspLSIsamp-1}, get
%	\begin{align}\label{eqn:PM-sampling-1} P(M)_{\delta^{\scriptsize\textrm{(spl)}}}{}&=\textrm{GFT}\, \textrm{diag}\left[\delta^{\scriptsize\textrm{(spl)}}\right]\textrm{GFT}^{-1} %=\textrm{GFT}\textrm{diag}\left[\textrm{(spl)}\right]\textrm{diag}\left[\textrm{(spl)}\right]\textrm{GFT}^{-1}
%	\end{align}
	%and $\widehat{s}=\left[\widehat{s}^T_K\widehat{s}^T_{N-K}\right]^T$.
%
\vspace*{-.5cm}
	\begin{align}\label{eqn:PM(K)-partition-1}
	    P_{\delta^{\scriptsize\textrm{(spl)}}}(M){}&\!\!=	\!\! \left[\!\!\!\begin{array}{cc}
	\textrm{GFT}_{K} \!\!\!&\!\!\! \textrm{GFT}_{N-K}
	\end{array}\!\!\!\right]\!
    \!\overbrace{\left[\!\!\!\begin{array}{cc}
    I_K\!\!&\!\!\\
    \!\!&\!\!0_{N-K}\end{array}\!\!\!\right]}^{\textrm{diag}\! \left[\!\delta^\textrm{(spl)}\!\right]}\!\!
 \left[\!\!\!\begin{array}{cc}
	\textrm{GFT}^{-1}_{K}\!\!\! &\!\!\! \textrm{GFT}^{-1}_{N-K}
	\end{array}\!\!\!\!\right]
\\
\label{eqn:PM(K)-partition-1-aa}
	{}&\!=\!\left[\!\begin{array}{cc}
	\textrm{GFT}_{K}\textrm{GFT}^{-1}_{KK}&\textrm{GFT}_{K}\textrm{GFT}^{-1}_{K(N-K)}
	\end{array}\!\right].
	\end{align}
Now, using the bandlimitedness of $\widehat{s}$ and $\delta^{\scriptsize\textrm{(spl)}}$ as in~\eqref{eqn:samplingdelta} in  equation~\eqref{eqn:gspvertexsampling-1} of result~\ref{res:gspsamplingasLSI-1}, we get
\begin{align}\label{eqn:deltaP(M)-1b}
 \hspace{-.3cm}\delta^{\scriptsize\textrm{(spl)}}\odot s{}& \!= \!\begin{bmatrix} s_K \\ 0 \end{bmatrix}\! \xrightarrow{\mathcal{F}} \! P_{\delta^{\scriptsize\textrm{(spl)}}}(M) \cdot \widehat{s}=\textrm{GFT}_{K}\textrm{GFT}^{-1}_{KK}\widehat{s}_K.
\end{align}
%From the left equation in~\eqref{eqn:deltaP(M)-1b}, it follows $s_d=s_K$.
Taking the $\textrm{GFT}$ of the left-hand side of~\eqref{eqn:deltaP(M)-1b}, we get
\begin{align} \label{eqn:sksmall}
     \hspace{-.5cm}\widehat{s}_\delta=\textrm{GFT}\begin{bmatrix}\! s_K \!\\ \!0 \! \end{bmatrix}{}&=P_{\delta^{\scriptsize\textrm{(spl)}}}(M) \cdot \widehat{s}\\
     \label{eqn:sksmall-2a}
     {}&=\textrm{GFT}_{K}\textrm{GFT}^{-1}_{KK}\widehat{s}_K\\
     \label{eqn:sksmall-3aa}
     {}&=\left[\begin{array}{c}
     \left[\textrm{GFT}_{K}\right]_{0K}\textrm{GFT}^{-1}_{KK}\\
     \cdots\\
     \left[\textrm{GFT}_{K}\right]_{iK}\textrm{GFT}^{-1}_{KK}\\
     \cdots\\
     \left[\textrm{GFT}_{K}\right]_{\left(\frac{N}{K}-1\right)K}\textrm{GFT}^{-1}_{KK}
     \end{array}\right]\widehat{s}_K,
     %
    %= P_{\delta^{\scriptsize\textrm{(spl)}}}(M)\widehat{s}
\end{align}
where we partitioned the $N\times K$ matrix $\textrm{GFT}_{K}$ into $\frac{N}{K}$ blocks $\left[\textrm{GFT}_{K}\right]_{iK}$, $i=0, \hdots, \frac{N}{K}-1$, where block $\left[\textrm{GFT}_{K}\right]_{iK}$ collects the~$K$ rows $iK, i\frac{N}{K}+1,\hdots, (i+1)K-1$. Then~\eqref{eqn:sksmall-3aa} shows that $\widehat{s}_\delta$ has $\frac{N}{K}$ copies $\left[\textrm{GFT}_{K}\right]_{iK}\textrm{GFT}^{-1}_{KK} \widehat{s}_K$ as we wanted to show.
Since each of the $\frac{N}{K}$ blocks of $\widehat{s}_\delta$ is a (filtered) replica of $\widehat{s}_K$ obtained by multiplying it with a $K\times K$ matrix block, no aliasing occurs.
      \end{proof}

Result~\ref{res:gspsamplingasLSI-1} and equation~\eqref{eqn:gspvertexsampling-1}, as well as result~\ref{res:GSPspectrumsubsampledsig} and equation~\eqref{eqn:sksmall-3aa}, show that, just like for DSP, GSP graph \textit{subsampling} has the dual interpretation of
\begin{inparaenum}[1)]
\item pointwise multiplication (modulation) $\delta^{\scriptsize\textrm{(spl)}}\odot s$ in the vertex domain; and
    \item LSI filtering in the spectral frequency domain. But further and very interestingly
    \item result~\ref{res:GSPspectrumsubsampledsig} and equation~\eqref{eqn:sksmall-3aa} show the spectrum \textit{replication} effect, with the spectrum of the sampled signal $\widehat{s}_\delta$ given by $\frac{N}{K}$ (filtered, possibly distorted) copies of the nonzero spectrum $\widehat{s}_K$ of~$s$. We note that we have assumed that we are sampling at rate~$K$ with no aliasing.
     \end{inparaenum}

     Although we have $\frac{N}{K}$ (filtered, possibly distorted) copies of $\widehat{s}_K$, we are not guaranteed that any of the blocks $\textrm{GFT}_{iK}$, $i=0, \hdots, \frac{N}{K}-1$ is full rank. This question will be taken care of when we consider decimation in section~\ref{subsec:decimation}.

       The next Theorem shows that $P(M)$ in result~\ref{res:gspsamplingasLSI-1} is the replicating filter equivalent to a train of frequency deltas when the GSP graph~$G$ is the directed cycle graph of DSP.

	\begin{theorem}[DSP $P(M)$]
	\label{thm:dftpm}
	Let: $G$ be a directed cycle graph of~$N$ nodes; $s$ a lowpass signal with cutoff frequency~$K$; $\textrm{I}_{K}$ the $K \times K$ identity matrix. Sampling with period $\frac{N}{K}$,\footnote{\label{ftn:stride}This samples uniformly every $\frac{N}{K}$, keeping~$K$ samples and zeroing $\frac{N}{K}-1$ samples in between.}
	\begin{align} \label{Peq}
	P_{\delta^{\scriptsize\textrm{(spl)}}}(M){}&\! =\!\frac{K}{N}\! \begin{bmatrix}
	\textrm{I}_{K} & \textrm{I}_{K} & \hdots & \textrm{I}_{K} \\
	\textrm{I}_{K} & \textrm{I}_{K} & \hdots & \textrm{I}_{K} \\
	\vdots  & \vdots&  \ddots & \vdots \\
	\textrm{I}_{K} & \textrm{I}_{K} & \hdots & \textrm{I}_{K} \\
	\end{bmatrix}\!.
	\end{align}
	\end{theorem}
	
A proof of Theorem \ref{thm:dftpm} can be found in the DSP literature \cite{Vaidyanathan}. Equation \eqref{eqn:gspLSIsamp-1} using $\text{GFT} = \text{DFT}$ and $\delta^{(spl)}$ as the uniform sampling also yields $P_{\delta^{\scriptsize\textrm{(spl)}}}(M)$.
From Theorem \ref{thm:dftpm}, we see that uniformly sampling in DSP produces a $P(M)$ that replicates exactly the band $\widehat{s}$. The spectrum of $\widehat{s}_\delta$ shows exact replications of~$\widehat{s}_K$ in DSP. For GSP and arbitrary graphs, by result~\ref{res:GSPspectrumsubsampledsig} and equation~\eqref{eqn:sksmall-3aa}, $\widehat{s}_\delta$  is also multiple replicated copies of~$\widehat{s}_K$, but the replicas may be distorted.

We may ask for which other graphs, besides the cycle graph, is the replicating filter LSI leading to $\widehat{s}_\delta$ to be $\frac{N}{K}$ exact replicas of~$\widehat{s}_K$. We provide two classes of graphs that, with specific choices of $\delta^{\scriptsize\textrm{(spl)}}$, also lead to exact replicating filters that are LSI filters, i.e., polynomials in~$M$.

Let $P_{\scriptsize\textrm{repl}}$ be the replicating LSI filter in Theorem~\ref{thm:dftpm}.

\begin{example}[Circulant graphs]\label{exp:circulantgraph} Interpret $P_{\text{repl}}$ as a circulant matrix. Its eigendecomposition is
\begin{align}\label{eqn:replicatingPrepl}
P_{\text{repl}} {}&= \text{DFT} \, \textrm{diag}\left[\delta^{\scriptsize\textrm{(spl)}}\right] \,\text{DFT}^{-1},
\end{align}
where $\delta^{\text{(spl)}}$ is uniformly sampling every $\frac{N}{K}$ values.

Consider that the graph is also given by circulant adjacency matrices but now with \text{distinct} eigenvalues. The graph Fourier transform is again $\text{GFT} = \text{DFT}$. Then, pre- and post-multiplying $P_{\text{repl}}$ in~\eqref{eqn:replicatingPrepl} by $\textrm{GFT}^{-1}$ and $\textrm{GFT}$
\begin{align}\label{eqn:expl-replicationfilter-1}
\text{GFT}^{-1}P_{\text{repl}} \text{GFT} = \text{diag}(\delta^{\text{(spl)}}),
\end{align}
 which is diagonal. The replicating filter $P_{\text{repl}}$ can then be written as a polynomial of~$M$ and is thus, LSI.

 We illustrate with the specific graph in figure~\ref{fig:circulant}. Its adjacency matrix~$A$ is circulant. In this example, every node of the graph is connected to its next node, its fourth next node and its sixth next node. By result~\ref{res:GSPshiftsAM},
$A = A_c + A_c^4 + A_c^6 = \text{DFT}^H \Lambda \text{DFT} = M$ where $A_c$ is the adjacency matrix of the $8 \times 8$ cyclic graph in \eqref{eqn:graphshiftA-1},  $\Lambda = \text{diag}(\lambda)$,
%      \begin{align}\label{eqn:cireigenvalues}
$\lambda = [3, -.29+,29j, -j, -1.7-1.7j, 1, -1.7, 1.7j, j, -.29,-.29j]^T$.
 %     \end{align}

      \begin{figure}[hbt]
    \centering
    \includegraphics[width=4.5cm,keepaspectratio]{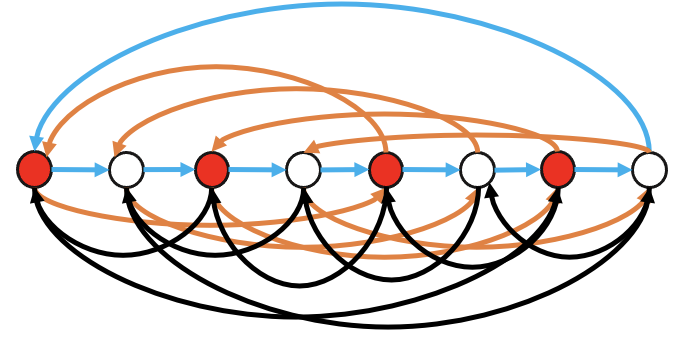}
    \vspace{-.2cm}
    \caption{Circulant Matrix: sampled nodes for replicating filter are red. Blue edges connect nodes with next nodes. Orange edges connect nodes with fourth next nodes. Black edges connect nodes with sixth next nodes.}
    \label{fig:circulant}
\end{figure}
 \end{example}
Let $K=4$. Then, we compute $\delta^{\scriptsize\textrm{(spl)}}$. As in equation~\eqref{eqn:expl-replicationfilter-1}:
$\textrm{diag}\left(\delta^{\scriptsize\textrm{(spl)}}\right) = \text{DFT}^H P_{\text{repl}}(M) \text{DFT} = [1,0,1,0,1,0,1,0]^T$.

We now compute the coefficients $p_{\scriptsize\textrm{repl}}$ of $P_{\scriptsize\textrm{repl}}(M)$ as a polynomial in~$M$.  Using \eqref{eqn:vertexsampling-LSIfilter-1v2} yields
$P_{\text{repl}}(M)\!\! =\!\! -.01 I_8 \!-\!.08M \!-\!.1M^2 \!+\! .18 M^3\! +\!.88 M^4 \!+\!.21M^5 \!-\!.03M^6\!-\!.05M^7\!$.

%Coefficients in $P_{\text{repl}}(M)$  rounded to~2 decimals places. We give a second example.
%
\begin{example}[Kronecker Product]\label{exp:KroneckerProduct}
Write $P_{\text{repl}}$ as
\begin{align}\label{eqn:KroneckerProduct}
P_{\text{repl}} = [11^T]_{\frac{N}{K}} \otimes I_K
\end{align}
 where $\otimes$ is the Kronecker product and $[1 1^T]_{\frac{N}{K}}$ is a $\frac{N}{K}\times \frac{N}{K}$ matrix of all 1s. This yields the eigendecomposition \begin{align}
 P_{\text{repl}} =\left(\text{DFT}_{\frac{N}{K}} \otimes V_K\right) \text{diag}\left([I_K^T, 0_{N-K}^T]^T\right) \left(\text{DFT}_{\frac{N}{K}}^{-1} \otimes V_K^{-1}\right),
 \end{align}
  where $V_K$ is any invertible matrix.

Consider graphs with adjacency matrix $A=A_{\frac{N}{K}} \otimes  B$ with unique eigenvalues where $A_{\frac{N}{K}}$ is the $\frac{N}{K}$ node cycle graph and~$B$ is any~$K$ node graph. The GFT of this graph is  $\text{DFT}_{\frac{N}{K}} \otimes V_K$ where $V_K$ is the GFT of~$B$. So:
\begin{align}
\text{GFT}^{-1}P_{\text{repl}} \text{GFT} =  \text{diag}(\delta^{\text{(spl)}})= \text{diag}\left([I_K^T, 0_{N-K}^T]^T\right)
\end{align}
 is diagonal.  The replicating filter $P_{\text{repl}}$ can be written as a polynomial of~$M$ and is thus, LSI.

We illustrate this example with the graph in figure~\ref{fig:kronecker} with adjacency
\begin{figure}[hbt]
    \centering
    \includegraphics[width=7.5cm,keepaspectratio]{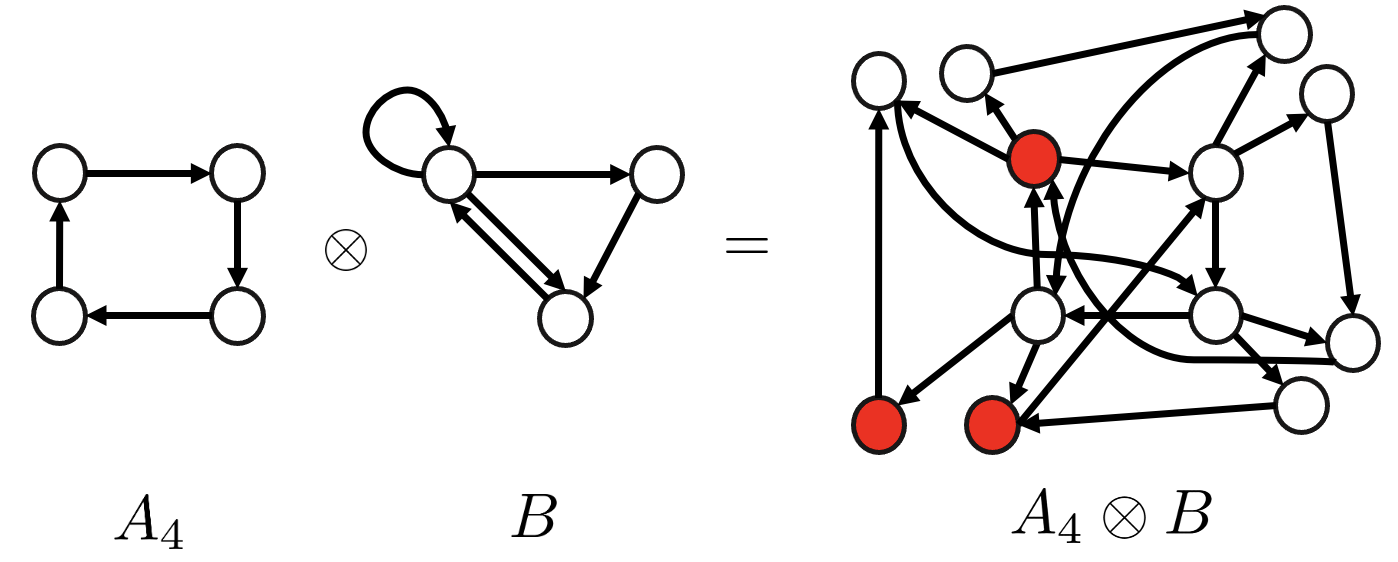}
    \caption{Kronecker Product Graphs. Red nodes are the sampled nodes for the replicating filter.}
    \label{fig:kronecker}
\end{figure}
matrix, $A \!= \!A_4 \!\otimes \!B$ where $A_4$ is the four node cycle graph and $B =\left[b_0\,b_1\,b_2\right]$ with $b_0=[0\,1\,0]^T$, $b_1=[1\,1\,1]^T$, and $b_2=[1\,0\,0]^T$. 
%$\begin{bmatrix}0&1&1\\1&1&0\\0&1&0\end{bmatrix}$. 
%Then, the $12 \times 12$ $A=A_4\otimes B$, where $A_4$ is the 4-dimensional cyclic shift.
%  $ \begin{bmatrix} 0 & 0 & 0 & B\\ B&0&0&0\\0&B&0&0\\0&0&B&0 \end{bmatrix}.$ 
 The eigendecompositions of $B$ and~$A$ are 
 \begin{align*}
 &B=\text{GFT}_B^{-1} \Lambda_B \text{GFT}_B\\
 &\Lambda_B= \text{diag}(\lambda_B),
     \lambda_B = [1.893,-.419+.606j,-.419-.606j]^T\\
&\text{GFT}_B^{-1}= \begin{bmatrix}
-.594&.676&.676\\-.707&-.402-.172j&-.402+.172j\\-.384&.119+.581j&.119-.581j
\end{bmatrix}\\
&A=\text{GFT}_A^{-1}\Lambda_A \text{GFT}_A, \text{GFT}_A = \text{DFT}_4 \otimes \text{GFT}_B\\
&\Lambda_A =\textrm{diag}\left[\lambda_B^T, -j\lambda_B^T, -\lambda_B^T, j\lambda_B^T\right].
\end{align*}
Then,
\begin{align*}
M {}&= \left(\text{DFT}_4 \otimes \text{GFT}_B\right) \Lambda_A^*  \left(\text{DFT}_4^{H} \otimes \text{GFT}_B^{-1}\right).
\end{align*}
Let $P_{\text{repl}}(M)$ be the replicating filter in Theorem \ref{thm:dftpm} with $K = 3$. We have
\begin{align*}
\text{diag}\left(\delta^{\text{(spl)}}\right) {}&\!\!= \!\!\text{GFT}_{A}^{-1} P_{\text{repl}}(M) \text{GFT}_{A}\!\! =\!\! [1,1,1,0,0,\hdots,0]^T.
\end{align*}
We compute LSI $P_{\text{repl}}(M)$. Using \eqref{eqn:vertexsampling-LSIfilter-1v2}, obtain
\begin{align*}
P_{\text{repl}}(M) {}&= .25I_{12}-.5M+.31M^2+.46M^3-1.38M^5\\
{}&+2.13M^6-.75M^7+.13M^9-.19M^{10}+.06M^{11}.
\end{align*}
\end{example}
\vspace*{-.4cm}
\subsection{Decimation}\label{subsec:decimation}
This section shows that vertex and spectral GSP decimation parallel DSP decimation. In DSP, decimation keeps the~$K$ sampled values and removes the $N-K$ zeros from the subsampled signal $s_{\delta}$. The~$N$ node cycle graph shrinks to the~$K$ node cycle graph with the $K \times K$ DFT in~\eqref{eqn:DFT1}. The signal spectrum ``stretches'' in frequency.
Likewise, in GSP, the decimated signal $s_d$ is the downsampled signal that keeps the~$K$ sampled values and removes the $N-K$ zeros. While the DSP ``decimated'' graph is the~$K$ node cycle graph, the GSP ``decimated'' graph $A_{d}$ is not as straightforward. We consider here the ``decimated'' signal~$s_d$, confirm the stretching of its graph spectrum $\widehat{s}_d$, and present the ``decimated'' $\textrm{GFT}_d$ and the ``decimated'' graph $A_{d}$.
\begin{result}[Decimated $s_d$ and $\textrm{GFT}_d$]\label{res:decimatedGFTd}
Let bandlimited $\widehat{s}=\left[\widehat{s}_K^T\,\,0_{N-k}^T\right]^T$ and  $\delta^{\scriptsize\textrm{(spl)}}$ as in~\eqref{eqn:samplingdelta}. Then,
\begin{align}
%\label{eqn:decimatedsd-1}
s_d{}&=s_K,
%\label{eqn:decimatedsd-1}
\widehat{s}_d=\widehat{s}_K,\textrm{   and   }
   \label{eqn:decimatedGFT-1}
    \textrm{GFT}_d=\textrm{GFT}_{KK},
\end{align} 
where $\textrm{GFT}_{KK}$ is the top left $K \times K$ subblock of the GFT (see equation~\eqref{eqn:partitionGFTcl-0}) and $\text{GFT}_d$ is the ``decimated'' GFT, the GFT of the ``decimated'' graph $A_d$.
\end{result}
\begin{proof}
From~\eqref{eqn:deltaP(M)-1b}, we have
\begin{align}\label{eqn:sksmall-3a}
 s_\delta{}&=\begin{bmatrix} s_K \\ 0 \end{bmatrix} =\textrm{GFT}^{-1}\textrm{GFT}_{K}\textrm{GFT}^{-1}_{KK}\widehat{s}_K.
 \end{align}
But
\vspace*{-.5cm}
\begin{align} \label{eqn:sksmall-2}
     \textrm{GFT}^{-1}\textrm{GFT}_{K}{}&=\left[\begin{array}{c}
     I_{KK}\\
     0_{(N-K)K}
     \end{array}\right],
\end{align}
since multiplication of the first~$K$ rows of $\textrm{GFT}^{-1}$ by $\textrm{GFT}_{K}$ gives $I_{KK}$ and the last $N-K$ rows of $\textrm{GFT}^{-1}$ are orthogonal to the columns in $\textrm{GFT}_{K}$. Substituting~\eqref{eqn:sksmall-2} in~\eqref{eqn:sksmall-3a}, get
\begin{align}\label{eqn:sksmall-3}
s_d{}&=s_K=\textrm{GFT}^{-1}_{KK}\widehat{s}_K.
\end{align}
Finally, we prove that $\textrm{GFT}^{-1}_{KK}$ is full rank and hence invertible. This follows because, by choice of the sampling set~$S$ (and sampling graph signal), by result~\ref{res:Sandsamplingdelta}, or equation~\eqref{eqn:vertexgraphsampling-1}, $s_K$ uniquely determines signal~$s$. By uniqueness of the $\textrm{GFT}$, $\widehat{s}=\left[\widehat{s}_K^T\,\,0_{N-k}^T\right]^T$ is uniquely determined from $s$ and hence from $s_K$. This also determines $\widehat{s}_K$ uniquely from $s_K$. Since~\eqref{eqn:sksmall-3} is a $K\times K$ linear relation between $s_K$ and $\widehat{s}_K$, we conclude that $\textrm{GFT}^{-1}_{KK}$ is full rank and thus invertible. Since $s_d=s_K$ and $\textrm{GFT}_d=\textrm{GFT}_{KK}$, we also get from~\eqref{eqn:sksmall-3} that $\widehat{s}_d=\widehat{s}_K$. The proof is complete.
\end{proof}
\vspace*{-.2cm}
Consider  graph $G_d$ with adjacency $A_d$ indexing $s_d$.
\begin{result}[Decimated graph $A_d$] \label{res:smallskfourier}
Let bandlimited $\widehat{s}=\left[\widehat{s}_K^T\,\,0_{N-k}^T\right]^T$ and the sampling signal $\delta^{\scriptsize\textrm{(spl)}}$ as in~\eqref{eqn:samplingdelta}, and $\textrm{GFT}_d=\textrm{GFT}_{KK}$. Then the decimated graph $A_d$ is
\begin{align} \label{eqn:smallskfourier}
A_d{}&=\text{GFT}_d^{-1}\cdot \Lambda_d \cdot \text{GFT}_d = \textrm{GFT}_{KK}^{-1}\cdot\Lambda_d \cdot\textrm{GFT}_{KK}
\end{align}
where $\Lambda_d=\textrm{diag}\left[\lambda_0\hdots \lambda_{K-1}\right]$.
\end{result}
\begin{proof}
We first determine the eigenvalues of the decimated graph. From~\eqref{eqn:Ps(M)froms}, the signal $s$ is a linear combination of the powers of the vector  $\lambda^*=\left[\lambda_0^*\, \lambda_1^* \hdots \lambda_{N-1}^*\right]^T$
\begin{align}\label{eqn:sintermslambdastar-1}
    s &= p_0 1 + p_1 \lambda^* +\hdots + p_{N-1} \lambda^{*(N-1)}.
    \end{align}
    Then, the sampled signal is
    \begin{align}\label{eqn:sintermslambdastar-2}
\hspace{-.5cm}\delta^{\scriptsize\textrm{(spl)}}\odot s &= \delta^{\scriptsize\textrm{(spl)}}\odot \left(p_0 1 + p_1 \lambda^* +\hdots + p_{N-1} \lambda^{*(N-1)}\right) \\
\label{eqn:eigenpowersampling}
&=  p_0 \!\left(\!\delta^{\scriptsize\textrm{(spl)}}\!\odot\! 1\!\right)\! + \!\hdots \!+\!p_{N-1} \! \left(\!\delta^{\scriptsize\textrm{(spl)}}\!\odot\!\lambda^{*(N-1)}\!\right)\!.
\end{align}
In \eqref{eqn:eigenpowersampling}, powers of vector $\lambda^n$ of eigenvalues are sampled by $\delta^{\scriptsize\textrm{(spl)}}$, zeroing out $N-K$ eigenvalues. Let $\lambda_d$ be the vector of non zeroed~$K$ eigenvalues (same ordering). They are the eigenvalues of $A_d$. Then~$A_d$ follows as in~\eqref{eqn:smallskfourier}.
\end{proof}
\begin{remark}[Sampling eigenvalues]\label{rem:samplingeigenvalues}
When sampling $s$ using $\delta^{\scriptsize\textrm{(spl)}}$, the eigenvalues are  sampled the same way. The chosen eigenvalues do not depend on which components of $\widehat{s}$ are zero, only on the choice of the sampling set.
\end{remark}
We finally consider the stretched spectrum of~$s_d$.
\begin{result}[GSP: Stretching]\label{res:gspstretching}
Given the set-up of the previous result, with~$s$ bandlimited, $\widehat{s}=\left[\widehat{s}_K^T\,0_{N-K}^T\right]^T$, the spectrum of the decimated signal $s_d$ is stretched over the full range of frequencies of the decimated graph~$A_d$.
\end{result}
\begin{proof}
By result~\ref{res:smallskfourier}, we see that the spectrum of the decimated signal~$s_d$  is $\widehat{s}_d=\widehat{s}_K$, and it occupies the full band $\Lambda_d$ of eigenvalues of the decimated graph $A_d$.
\end{proof}
Note that $\Lambda_d$ is a subset of eigenvalues of~$A$, so stretching has the same interpretation in both DSP and GSP.

The next example illustrates how to derive the DSP vertex (time) and frequency interpretations for $s_d$ when the graph is the time directed cycle graph.
\begin{example}[DSP Example]\label{exp:dspexample-1}
Sample uniformly $s$ defined on cycle graph with adjacency $A$, taking every \!$\frac{N}{K}$\! samples. Let decimated signal be $s_d\!=\!s_K$ and  $\widehat{s} \!=\! \left[\!\widehat{s}_K^T, 0^T\!\right]^T$\!\!.

The eigenvalues of $A$ are $\lambda_k=e^{-j\frac{2\pi}{N}k}$. Sampling uniformly produces $\lambda_{d,r}=e^{-j\frac{2\pi}{N}\left(\frac{N}{K}\right)r} = e^{-j\frac{2\pi}{K}r}$, $r= 0,1,\hdots,K-1$. These are the eigenvalues of the~$K$ node cycle graph $C_K$ illustrating the need to sample uniformly. Not sampling uniformly does not choose the eigenvalues of $C_K$. Finally, the decimated graph is $A_{d} \!\!=\!\! \textrm{DFT}_d^{-1}\! \cdot\!\Lambda_{d} \! \cdot\!\textrm{DFT}_d$, the~\!$K$\! node cycle graph \!$C_K$ where $\text{DFT}_d$ is the $K\times K$ DFT in \eqref{eqn:DFT1}.

This shows again for DSP that the spectrum of the subsampled signal $s_\delta$ is $\frac{N}{K}$ replicas of the low pass spectrum $\widehat{s}_K$. Using \eqref{eqn:sksmall-3} and Theorem \ref{thm:dftpm}
\begin{align} \label{eqn:dspexsk}
P_{\delta^{\scriptsize\textrm{(spl)}}}(M) \widehat{s}{}&=\text{GFT}_K s_K=
\begin{bmatrix}
\text{DFT}_{d}\\\vdots\\ \text{DFT}_{d}
\end{bmatrix} s_K = \begin{bmatrix}
\widehat{s}_K\\\vdots\\ \widehat{s}_K
\end{bmatrix},
\end{align}
%
%Note that in this example, we assume no aliasing. However, if the signal were aliased, the signal would still be $K$ periodic (just perhaps, not with replications of $\widehat{s}_K$) and we would still proceed as follows. Of course, the aliased signal is not recoverable, but we can still decimate.
%
%Choosing the first $K$ rows of \eqref{eqn:dspexsk} and applying \eqref{eqn:smallskfourier} yields $\text{DFT}_s s_K = \widehat{s_K}$ with $\text{GFT}_{sm} = \text{DFT}_s$. This equation is the Fourier relationship after decimiating in DSP. The Fourier transform of the decimated signal $s_d = s_K$ is $\widehat{s}_K$, the first $K$ values of $P(M)\widehat{s}$.
%
%Instead, if we were to choose any $K$ linearly independent rows of~\eqref{eqn:dspexsk}, since we must choose all $K$ rows of the $\text{DFT}_d$ (in some order), we obtain $P \text{DFT}_s s_K = P \widehat{s}_K$ for some permutation matrix $P$. In this case, the smaller graph is $A_{sm}' = \text{DFT}_s^{-1} P^{-1} \Lambda_{sm} P \text{DFT}_s$.
\end{example}
%
%\textbf{Jose: NOTE: This has to do with the fact that we are keeping $\Lambda_{sm}$ and $s_K$ fixed. We reordered the frequencies here, but not the eigenvalues. If we started with $M_{sm}' = P M_{sm} P^{-1}$, we see that $A_{sm}' = P A_{sm} P^{-1}$ since $A=M$. This is a similar issue to Reviewer 2's remarks and I think we can either address it or cut out the last paragraph. }

\begin{remark}[DSP: Stretching]\label{rmk:DSPstretching}
The original signal is at~$N$ frequencies $2\pi l/N$, $l = 0,1,\hdots,N-1$, while the decimated signal is at~$K$ frequencies, $2\pi l/K$, $l = 0,1,\hdots,K-1$. This ``stretches'' the signal spectrum to fill the $2\pi$ range.
\end{remark}
%\begin{example}[TBD: Strong product of graphs, sampling]
%\end{example}

\section{GSP Sampling: Upsampling and Interpolation}\label{sec:upsamplinginterpolation}
We assume the same set-up as described in the introduction to section~\ref{sec:subsamplingdecimation}. We start in subsection~\ref{subsec:reconstructionupsampling} with upsampling. Then in subsection~\ref{subsec:perfectreconstruction} we address the conditions of when does a given sampling signal $\delta^{\scriptsize\textrm{(spl)}}$ lead to perfect reconstruction, and finally in subsection~\ref{subsec:reconstructionasfiltering} we explore GSP interpolation as filtering.

\subsection{Reconstruction: Upsampling}\label{subsec:reconstructionupsampling}
Upsampling in DSP reintroduces the zeros into the $K \times 1$ signal, $s_d$, producing the sampled $N \times 1$, $s_{\delta}$. The $K$ node cycle graph becomes the $N$ node cycle graph. We emphasize that in DSP we know:
\begin{inparaenum}[1)]
\item the larger and decimated graphs  ($N$ and~$K$ node cycle graphs) and their adjacency matrices~$A$ and~$A_d$;
    \item the positions of the zeros when adding the zeros back into $s_d$;
     \item the eigenvalues of~$A$ and~$A_d$; and
      \item the $\textrm{DFT}_N$ and $\textrm{DFT}_K$.
       \end{inparaenum}

     In GSP, to upsample, we also need to know
      \begin{inparaenum}[1)]
      \item both the original and downsampled graphs~$G$ and $G_d$ and their adjacency matrices~$A$ and~$A_d$;
          \item the positions of the zeros when adding the zeros back into $s_d$;
          \item $\Lambda$ and $\Lambda_d$; and
          \item $\textrm{GFT}$ and $\textrm{GFT}_d$.
           \end{inparaenum}
           Then, upsampling starts with padding zeros to $s_d$ to produce $s_\delta$ in the vertex domain. In the spectral domain, we obtain $P(M) \widehat{s}$. Now, $\widehat{s}$ no longer extends over the frequency range of~$A$ since it is bandlimited.
     %This allows us to return to \eqref{eqn:gspvertexsampling-1} and the subsampled signal $s_\delta$.

\subsection{Reconstruction: Perfect reconstruction}\label{subsec:perfectreconstruction}% Condition on $\delta^{\scriptsize\textrm{(spl)}}$}
\label{subsec:gspreconstruction}
%In DSP, Shannon's theorem \cite{oppenheimwillsky-1983,siebert-1986,oppenheimschaffer-1989,mitra-1998} reconstructs perfectly a band-limited signal, $s_r=s$, by ideal filtering\textemdash windowing the spectrum of the decimated signal $s_d$ or interpolating (convolving) the upsampled signal $s_\delta$ with a periodic train of (discrete) $\textrm{sinc}$ functions. We consider GSP reconstruction of a decimated band-limited signal.

Let $s\xleftrightarrow{\mathcal{F}} \widehat{s}$. Assume~$s$ has bandwidth~$K$, i.e., $\left\|\widehat{s}\right\|_0\leq K$, and that after possible permutation $\widehat{s}=\left[\widehat{s}^T_K \widehat{s}^T_{N-K}\right]^T$ with $\widehat{s}_{N-K}=0$. By result~\ref{res:Sandsamplingdelta} there is a sampling set~$S$ with characteristic function $\delta^{\scriptsize\textrm{(spl)}}$ such that perfect reconstruction is possible from the samples $s_d$ in~$S$, i.e., $s=s_r$, with $s_r$ reconstructed from $s_d$, for example, using \eqref{eqn:vertexgraphsampling-1}.

Now we address a different question for bandlimited~$s$ with bandwidth$=K$. Let the sampling signal $\delta^{\scriptsize\textrm{(spl)}}$ have $\left\|\delta^{\scriptsize\textrm{(spl)}}\right\|=K$. The question is when does $\delta^{\scriptsize\textrm{(spl)}}$ lead to perfect reconstruction of~$s$. In other words, when can we recover~$s$ from $s_d=s_K$ where $s_K$ is obtained by discarding the zeros of the subsampled $s_\delta=\delta^{\scriptsize\textrm{(spl)}}\odot s$. 

Let $\delta^{\scriptsize\textrm{(spl)}}\!\!=\!\!\left[\!1_K^T 0^T_{N-K}\!\right]^T$ and recall $P_{\delta^{\scriptsize\textrm{(spl)}}}(M)$ in~\eqref{eqn:partitionGFTcl-4}. Let $P_{\delta^{\scriptsize\textrm{(spl)}}}(M)_K$ be its first~$K$ columns in its partition~\eqref{eqn:PM(K)-partition-1-aa}.
%This turns out to be based on a rank condition of a submatrix of the spectral LSI filter $P_{\delta^{\scriptsize\textrm{(spl)}}}(M)$.
%
%
%
\begin{result}[Rank of $P_{\delta^{\scriptsize\textrm{(spl)}}}(M)_K$\!\!\! ]\label{res:P(M)_Kfullrank}  Then
\begin{align}\label{eqn:P(M)_Kfullrank-1}
\textrm{rank}\left[P_{\delta^{\scriptsize\textrm{(spl)}}}(M)_K\right]= \textrm{rank}\left(\textrm{GFT}_K\textrm{GFT}^{-1}_{KK}\right)=K
\end{align}
 iff the $K\times K$ square  matrix  $\textrm{GFT}^{-1}_{KK}$ is  invertible.
\end{result}	
\begin{proof}
\textit{Only if}: We have $\textrm{rank}\left(\textrm{GFT}_K\right)=K$ and 
%\begin{align}\label{eqn:P(M)_Kfullrank-2}
   %\hspace{-.4cm}
  %\hspace{-.3cm} 
  $
  \textrm{rank}\left(P_{\delta^{\scriptsize\textrm{(spl)}}}(M)_K\right)\leq \min\left(\textrm{rank}\left(\textrm{GFT}_K\right),\textrm{rank} \left(\textrm{GFT}^{-1}_{KK}\right)\right)$. 
%\end{align}
Then, if $\textrm{GFT}^{-1}_{KK}$ not invertible, its rank $\!\!\!<\!\!\!K$, and $\textrm{rank}\!\left(\!P_{\delta^{\scriptsize\textrm{(spl)}}}(M)_K\!\right)\!<\! K$.

\textit{If}: If $\textrm{rank}\left(\textrm{GFT}^{-1}_{KK}\right)=K\Longrightarrow \textrm{rank}\left(P_{\delta^{\scriptsize\textrm{(spl)}}}(M)_K\right)= \textrm{rank}\left(\textrm{GFT}_K\right)=K$.
\end{proof}
%
%Without loss of generality, let $\widehat{s}=\left[\widehat{s}_K^T\,\widehat{s}^T_{N-K}\right]^T$.
\begin{result}[Perfect reconstruction sampling  condition]\label{res:perfectreconstruction-a1}
Without loss of generality, let $\delta^\textrm{(spl)}=\left[1_K^T 0^T_{N-K}\right]^T$. Assume $\widehat{s}_{N-K}=0$. The signal~$s$ can be perfectly reconstructed from  $s_d\!=\!s_K$ iff $\textrm{rank}\left(\textrm{GFT}^{-1}_{KK}\right)\!=\!K$.
\end{result}
\begin{proof}
With~$s$ lowpass,
\begin{align}\label{eqn:perfectreconstructioncondition-1}
\hspace{-.4cm}\delta^{\scriptsize\textrm{(spl)}}\odot s{}&=\delta^{\scriptsize\textrm{(spl)}}\odot \left[\begin{array}{c}
s_K\\
s_{N-K}
\end{array}\right]=\left[\begin{array}{c}
s_K\\
0_{N-K}
\end{array}\right]\\
\label{eqn:perfectreconstructioncondition-2}
{}&=\textrm{GFT}^{-1}P_{\delta^{\scriptsize\textrm{(spl)}}}(M)_K \widehat{s}_K\\
\label{eqn:perfectreconstructioncondition-3}{}&=\textrm{GFT}^{-1}\textrm{GFT}_K \textrm{GFT}^{-1}_{KK}\widehat{s}_K\\
\label{eqn:perfectreconstructioncondition-4}{}&=\left[\begin{array}{c}
\textrm{GFT}^{-1}_{KK}\\
0_{N-K}
\end{array}\right]\widehat{s}_K.
\end{align}
Given $s_d=s_K$, $\widehat{s}_K$ is determined from~\eqref{eqn:perfectreconstructioncondition-4} iff $\textrm{GFT}^{-1}_{KK}$ is invertible, from which~$s$ is perfectly reconstructed.
\end{proof}
This result seems repetitive when contrasted with result~\ref{res:decimatedGFTd} and equation~\eqref{eqn:sksmall-3}. The difference is that in result~\ref{res:decimatedGFTd} and equation~\eqref{eqn:sksmall-3} we assume that $\delta^{\scriptsize\textrm{(spl)}}$ corresponds to a sampling set~$S$ for which we know we can reconstruct perfectly~$s$ from $s_d$, while here we are given a $\delta^{\scriptsize\textrm{(spl)}}$ and have to find conditions for perfect reconstruction of~$s$ from $s_d$.

Result~\ref{res:perfectreconstruction-a1} provides how to reconstruct in the spectral domain, shown in result \ref{res:reconstructionspdomain-1}.
\begin{result}[Reconstruction in spectral domain]\label{res:reconstructionspdomain-1} Under result~\ref{res:perfectreconstruction-a1} assumptions, let  bandlimited~$s$ with bandwidth~$K$ be  decimated to $s_d=s_K$. Then $s$ is reconstructed by
\begin{align}\label{eqn:reconstructionspdomain-1}
    \widehat{s}_K=\left[\textrm{GFT}^{-1}_{(KK)}\right]^{-1} s_K \Longrightarrow
    s&=\textrm{GFT}^{-1}\left[\begin{array}{c}
    \widehat{s}_K\\
    0_{N-K}
    \end{array}\right].
\end{align}
\end{result}
 The proof follows from result~\ref{res:perfectreconstruction-a1} and~\eqref{eqn:perfectreconstructioncondition-4}.

 Partitioning $\textrm{GFT}^{-1}$ as in~\eqref{eqn:partitionGFTcl-3}, get from result~\ref{res:reconstructionspdomain-1}
\begin{align}
 \label{eqn:reconstructionspdomain-2}
   s &=\left[\begin{array}{c}
    I_K\\
    \textrm{GFT}^{-1}_{(N-K)K}\left[\textrm{GFT}^{-1}_{(KK)}\right]^{-1}\\
    \end{array}\right]s_K.
\end{align}
This shows it is possible to recover $s$ from $s_d=s_K$. But, like for DSP and Shannon reconstruction, it is important to find equivalent filtering interpretations for reconstruction, in both the vertex and the spectral domains. The next subsection explores this.
\subsection{Reconstruction: Interpolation as Filtering} \label{subsec:reconstructionasfiltering}
Result~\ref{res:reconstructionspdomain-1} shows one way to reconstruct~$s$ from~$s_d=s_K$. In DSP, Shannon's Sampling Theorem reconstructs the signal by ideal lowpass \textit{filtering} the upsampled signal. Likewise, we consider a GSP \textit{filtering} approach  to reconstruct~$s$ from the upsampled $s_\delta$. This parallels section~\ref{subsec:lsisampling}, where downsampling is by \textit{LSI spectral} filtering, see~\eqref{eqn:gspvertexsampling-1}, result~\ref{res:gspsamplingasLSI-1}. We show that reconstruction from an upsampled signal can be achieved by spectral domain filtering, but, in contrast with section~\ref{subsec:lsisampling}, the reconstruction filter is not in general LSI.
\vspace*{-.2cm}
\begin{result}[Reconstruction by filtering]\label{res:reconstructionbyfiltering} Let~$s$ be bandlimited with bandwidth~$K$, $\widehat{s}=\left[\widehat{s}_K^T\,0_{N-K}^T\right]^T$. Let $\delta^{\scriptsize\textrm{(spl)}}=\left[1_K^T 0^T_{N-K}\right]^T$ be the sampling signal for sampling set~$S$, and~$s$ be decimated to $s_d=s_K$ by $\delta^{\scriptsize\textrm{(spl)}}$. Then reconstruct $s$  by filtering upsampled $s_\delta$ as follows:
\begin{align}\label{eqn:reconstructionbyfiltering-01}
s{}&=P_{\delta^{\scriptsize\textrm{(spl)}}}(A)\cdot F\cdot s_\delta\xrightarrow{\mathcal{F}}\left[\begin{array}{c}
1_K\\
0_{N-K}
\end{array}\right]\odot Q\cdot\widehat{s}_\delta
\end{align}
where\vspace*{-.5cm}
\begin{align}\label{eqn:reconstructionbyfiltering-02}
% \hspace{-.5cm}
 &Q=
 \begin{blockarray}{c}
 \begin{block}{[c]}
Q_{KK}\hspace{.5cm}Q_{K(N-K)}\\
\BAhhline{.}
%\cdots&\cdots&\cdots\\
\hspace{-.2cm}Q_{(N-K)N}\\
\end{block}
\end{blockarray},
\hspace{.2cm}
F{}=\textrm{GFT}^{-1}\cdot Q\cdot\textrm{GFT}\\
\label{eqn:reconstructionbyfiltering-03}
&Q\cdot P_{\delta^{\scriptsize\textrm{(spl)}}}(M)=\begin{blockarray}{c}
 \begin{block}{[c]}
I_{KK}\hspace{.7cm}B_{K(N-K)}\\
\BAhhline{.}
\hspace{-.2cm}B_{(N-K)N}\\
\end{block}
\end{blockarray}\\
\label{eqn:reconstructionbyfiltering-05}
&P(A)=\textrm{GFT}^{-1}\left[\!\!\!
\renewcommand*{\arraystretch}{.4}\begin{array}{cc}
I_{KK}\!\!&\!\!\\
\!\!&\!\!0_{(N-K)(N-K)}
\end{array}\!\!\!\right]\!\textrm{GFT}
\end{align}
where $B_{K(N-K)}$ and $B_{(N-K)N}$ are non prescribed.
\end{result}
We interpret the result before proving it. The left side of~\eqref{eqn:reconstructionbyfiltering-01} reconstructs~$s$ by filtering in the vertex domain the upsampled $s_\delta$. The right-hand side, reconstructs~$\widehat{s}$ by filtering in the spectral domain the upsampled $\widehat{s}_\delta$. Equation~\eqref{eqn:reconstructionbyfiltering-01} gives GSP reconstruction in the spectral domain as filtering with filter~$Q$ followed by lowpass LSI ideal filtering (Hadamard product with $\left[1_K^T\,0_{N-K}^T\right]^T$). In the vertex domain (left side of~\eqref{eqn:reconstructionbyfiltering-01}), we have first the non LSI filtering by~$F$, followed by the ideal lowpass LSI filter $P(A)$, whose frequency response is $\left[1_K^T\,0_{N-K}^T\right]^T$. In DSP, as we will show below, $Q$ and~$F$ are trivial filters, and reconstruction is limited to ideal lowpass $P(A)$ (impulse response is the discrete sinc) with flat frequency response over the signal band. In~\eqref{eqn:reconstructionbyfiltering-01}, the reconstruction filter~$Q$ is not necessarily LSI, i.e., not necessarily polynomial in the spectral shift~$M$. Certain blocks like $Q_{(N-K)N}$ are not constrained. Its equivalent in the vertex domain is filter~$F$ that again is not, in general, LSI, i.e., not polynomial in~$A$. On the other hand, $\left[\begin{array}{c}
1_K\\
0_{N-K}
\end{array}\right]$, is an ideal lowpass LSI filter whose vertex equivalent is the LSI polynomial filter $P(A)$.
\begin{proof}
By result~\ref{res:Ps(M)directfroms} and~\eqref{eqn:Ps(M)froms-2} and graph signal  $\delta^{\scriptsize\textrm{(spl)}}$
\begin{align}\label{eqn:P_deltaspl(M)-1}
P_{\delta^{\scriptsize\textrm{(spl)}}}(M)= \textrm{GFT}\textrm{diag}\left[\delta^{\scriptsize\textrm{(spl)}}\right]\textrm{GFT}^{-1}.
\end{align}
Since $\left\|\delta^{\scriptsize\textrm{(spl)}}\right\|_0=K$, $\textrm{rank}\left(P_{\delta^{\scriptsize\textrm{(spl)}}}\right)=K$.
From~\eqref{eqn:gspvertexsampling-1}, result~\ref{res:gspsamplingasLSI-1},
\begin{align}\label{eqn:P_deltaspl(M)-1a}
P_{\delta^{\scriptsize\textrm{(spl)}}}(M)\left[\begin{array}{c}
\widehat{s}_K\\
0_{N-K}
\end{array}\right]{}&=\widehat{s}_\delta.
\end{align}
Gauss Jordan elimination (GJ-E) determines a~$Q$ such that $P_{\delta^{\scriptsize\textrm{(spl)}}}(M)$ is in reduced row echelon form. After GJ-E, \eqref{eqn:P_deltaspl(M)-1a} becomes
\vspace*{-.2cm}
\begin{align}\label{eqn:P_deltaspl(M)-2}
Q\cdot P_{\delta^{\scriptsize\textrm{(spl)}}}(M)\left[\begin{array}{c}
\widehat{s}_K\\
0_{N-K}
\end{array}\right]{}&=Q\cdot\widehat{s}_\delta
\\
\label{eqn:P_deltaspl(M)-3}
\Longrightarrow\raisebox{-.17cm}{$\begin{blockarray}{c}
 \begin{block}{[c]}
I_{KK}\hspace{.7cm}\widetilde{P}_{K(N-K)}\\
\BAhhline{.}
\hspace{-.3cm}0_{(N-K)N}\\
\end{block}
\end{blockarray}$}\left[\begin{array}{c}
\widehat{s}_K\\
0_{N-K}
\end{array}\right]{}&=\left[\begin{array}{c}
\widehat{s}_{\delta_K}\\
0_{N-K}
\end{array}\right]
\vspace*{-.4cm}
\end{align}
from which $\widehat{s}_{K}=\widehat{s}_{\delta_K}$. The ideal lowpass filter $\left[\!\!\begin{array}{cc}
1_K^T&
0_{N-K}^T
\end{array}\!\!\right]^T$ still recovers $\left[\!\!\begin{array}{cc}
\widehat{s}_K^T&
0_{N-K}^T
\end{array}\!\!\right]^T$ whose $\textrm{GFT}^{-1}$ reconstructs~$s$.

The left-hand side in~\eqref{eqn:reconstructionbyfiltering-01} follows since spectral filtering by~$Q$ is in the vertex domain filtering by~$F$ given in~\eqref{eqn:reconstructionbyfiltering-02}, and ideal lowpass filtering in the spectral domain is in the vertex domain LSI filtering by $P(A)$ given in~\eqref{eqn:reconstructionbyfiltering-05}.
\end{proof}
\begin{remark}\label{rem:reconsctructionbyfiltering}
\begin{inparaenum}[1)]
\item The~$Q$ obtained by GJ-E in the proof of result~\ref{res:reconstructionbyfiltering} is not unique, and there are other methods to design a  spectral filter~$Q$ such that $Q\cdot P_{\delta^{\scriptsize\textrm{(spl)}}}(M)$ has the block form in~\eqref{eqn:reconstructionbyfiltering-03}.
\item If there is column swapping in GJ-E, then one needs to account for a permutation~$\Pi$.
\item In the block form of~\eqref{eqn:reconstructionbyfiltering-03}, only the $I_{KK}$ block matters, since the zero block in the lowpass signal~$\widehat{s}$ multiplies   $\widetilde{P}_{K(N-K)}$ and the ideal lowpass filter filters out $\widetilde{P}_{(N-K)N}$. This provides degrees of freedom in designing spectral filter~$Q$.
\item By row and column permutation, we can rearrange~$P_{\delta^{\scriptsize\textrm{(spl)}}}$ so that its block $P_{\delta^{\scriptsize\textrm{(spl)}}}(M)_{KK}$ is invertible. Then,
\begin{align}\label{eqn:aQthatworks-1}
 Q{}&=
 \begin{blockarray}{c}
 \begin{block}{[c]}
\left[P_{\delta^{\scriptsize\textrm{(spl)}}}(M)_{KK}\right]^{-1}\hspace{.5cm}0_{K(N-K)}\\
\BAhhline{.}
%\cdots&\cdots&\cdots\\
\hspace{-.2cm}Q_{(N-K)N}\\
\end{block}
\end{blockarray}
\end{align}
can be used, with $Q_{(N-K)N}$ designed to possibly achieve other design considerations.
\item Result~\ref{res:reconstructionbyfiltering} \framebox{reconstructs~$s$ by filtering} from its decimated~$s_d$ and upsampled $s_\delta$ versions, paralleling the DSP uniform  sampling reconstruction by ideal lowpass filtering. In general~$Q$ is not LSI, i.e., a polynomial in~$M$. Given the degrees of freedom in~$Q$, one can in some cases find a LSI version of~$Q$, in which case reconstruction in GSP is, like in DSP, achieved by LSI filtering, see next example.
\end{inparaenum}
\end{remark}
\begin{example}\label{exp:LSIsamplingreconstruction-1}
\label{subsec:GSPgeneralexample}
Consider the five node star graph  in Example~\ref{exp:stargraphM} with bandlimited ($K = 2$) $s = [-2\,3\,3\,3\,3]^T$, $\widehat{s} = [1\,2\,0\,0\,0]^T$. Let $\delta^{\scriptsize\textrm{(spl)}} = [1\,1\,0\,0\,0]^T$.
By~\eqref{eqn:P_deltaspl(M)-1a},
\begin{align}  \label{eqn:pmexamplestar}
   & P_{\delta^{\scriptsize\textrm{(spl)}}}(M) {}= \frac{1}{4}M^2
    = \left[\begin{array}{cc}
    B_{11}&B_{12}\\
    B_{12}^T&B_{22}\end{array}\right]\\
    %B_{11}{}&= \left[\renewcommand{\arraystretch}{.5}\!\!\begin{array}{cc}
    %.625\!\!&\!\!-.375\\
    %-.375\!\!&\!\!.675\end{array}\!\!\right]\!, B_{12}=.177\left[\!\!\begin{array}{c}
    %1_3^T\\
    %1_3^T\end{array}\!\!\right]\!,
   & B_{11}{}= \left[\begin{array}{cc}
    .625&-.375\\
    -.375&.675\end{array}\right], B_{12}=.177\left[\begin{array}{c}
    1_3^T\\
    \\
    1_3^T\end{array}\right], B_{22}=.25I_3.
\end{align}
Since $B_{11}$ is invertible, a possible filter is  $Q=\textrm{blockdiag}\left[
Q_{11},4I_3\right]$, with $Q_{11}=\left[q_1\,q_2\right]$, $q_1=\left[2.5\,1.5\right]^T$ and $q_2=\left[1.5\,2.5\right]^T$. This Q is  LSI, $Q=4I_5 - .75 M - .375 M^2$.

The ideal lowpass filter  $\left[1_2^T\,0^T_3\!\right]$ is in the vertex domain $P(A)\!=\!\frac{1}{4}A^2$\!.

Reconstruction: Signal~$s$ is reconstructed from $s_\delta$ and  $\widehat{s}_\delta$ in the vertex and spectral domains, both by \textbf{LSI filtering}:
\vspace*{-.3cm}
%{{\scriptsize
\begin{equation}\label{eqn:examplefullpicture}
\underbrace{\textcolor{red}{\frac{1}{4}A^2}}_{\substack{\text{l.p.f.} \\ P(A) \\ \text{in}\\ \text{vertex}}}\hspace{-.3cm}
\underbrace{\textcolor{orange}{\begin{bmatrix}1\\4\\4\\4\\4\end{bmatrix} }}_{\substack{F=Q\\ \text{ in vertex}}}\hspace{-.5cm}\odot\hspace{-.1cm} \underbrace{\textcolor{blue}{\overbrace{\begin{bmatrix}1\\1\\0\\0\\0\end{bmatrix}}^{\delta^{\scriptsize\textrm{(spl)}}} \hspace{-.2cm}\odot\hspace{-.1cm}} \overbrace{\begin{bmatrix} -2\\3\\3\\3\\3\end{bmatrix}}^{s}}_{s_\delta} \hspace{-.2cm} \xrightarrow{\mathcal{F}}\hspace{-.3cm}
\underbrace{\textcolor{red}{\begin{bmatrix}1\\1\\ 0 \\0\\0 \end{bmatrix} }}_{\text{\textrm{l.p.f.}}}
\hspace{-.2cm}\odot \textcolor{orange}{\overbrace{\left(\hspace{-.05cm}4I -\hspace{-.1cm}.75 M -\hspace{-.1cm}.375 M^2\hspace{-.05cm}\right)}^{Q}}\hspace{-.1cm} \underbrace{\textcolor{blue}{\overbrace{\left(\frac{M^2}{4}\right)}^{P(M)}} \hspace{-.2cm}\overbrace{\begin{bmatrix}1\\2\\0\\0\\0\end{bmatrix}}^{\widehat{s}}}_{\widehat{s}_\delta}
\end{equation}%%
%}}
%\vspace*{-.1cm}
\end{example}
\vspace*{-.55cm}

\subsection{DSP from GSP: Nyquist-Shannon Sampling}
\label{subsec:nyquistshannon}
To illustrate the impact of different GSP choices, we consider a simple example when $G$ is the cycle graph and with DSP Nyquist-Shannon sampling. Let $N = 4$, signal~$s$ with four values, bandlimited with $K=2$, and sampled uniformly. $A$ is the directed cycle graph of 4 nodes. As shown in section~\ref{sec:spectralshift}, $A = M$. Also, in DSP, since the eigenvalues are all unique, multiplication in one domain is filtering in the other with a polynomial filter, either $\textrm{P}(A)$ or $\textrm{P}(M)$.

Nyquist-Shannon sampling  is illustrated by:
\begin{equation} \label{eqn:recoverynyquist}
\underbrace{\textcolor{red}{\text{P}(A)}}_{\text{sinc}} \overbrace{\underbrace{\textcolor{blue}{\begin{bmatrix}
    1\\ 0\\1\\0
    \end{bmatrix}\odot}}_{\delta^{\scriptsize\textrm{(spl)}}}  \underbrace{\begin{bmatrix}
    s_0\\s_1\\s_2\\s_3
    \end{bmatrix}}_{s}}^{s_\delta} \xrightarrow{\mathcal{F}} \underbrace{\textcolor{red}{\begin{bmatrix}
   2\\2\\0\\0 \end{bmatrix}\odot}}_{\text{l.p.f.}} \overbrace{\underbrace{\textcolor{blue}{\frac{1}{2}
    \begin{bmatrix}
    I_2 & I_2 \\ I_2 & I_2
    \end{bmatrix}}}_{\textrm{P}(M)}
    \underbrace{\begin{bmatrix}
    \widehat{s}_0\\ \widehat{s}_1\\0\\0
    \end{bmatrix}}_{\widehat{s}}}^{\widehat{s}_s}
\end{equation}
where $I_2$ is the $2-dimensional$ identity. Colors indicate same operation in both domains.

 We show how Nyquist-Shannon recovery derives from GSP sampling. Let $Q_2$ be the upper half of~$Q$. The condition on filter~$Q$ is for its upper half to satisfy
\begin{equation} \label{conditionexdsp}
    Q_2 \textrm{P}(M)_2 = Q_2 \frac{1}{2}  \begin{bmatrix}
    I_2 \\ I_2
    \end{bmatrix} = I_2
\end{equation}
where filters are indexed by their dimension. We consider three different GSP choices for $Q_2$ that satisfy~\eqref{conditionexdsp}.
\begin{s_enumerate-0}
    \item $Q_2 = 2 \begin{bmatrix} I_2 & 0\\ \end{bmatrix}$. We can fill in the bottom $N-K$ rows of $Q$ to produce a $Q$ that is LSI:\footnote{An LSI filter in the time domain is $P(A)$ and in the frequency domain is $P(M)$. Since $A$ is the directed cycle graph adjacency matrix and $M = A$, all LSI filters in DSP are circulant matrices.}
    $Q=2 \textrm{blockdiag}\left[I_2,I_2\right]= 2 I_4$. This~$Q$ is equivalent to multiplying $s_\delta$ by $2 [1,1,1,1]^T$ in the time domain. Thus, $Q$ yields the following recovery:
    \vspace*{-.2cm}
    %{{\scriptsize
    \begin{equation} \label{recoveryq1}
%\underbrace{\textcolor{red}{\text{P}(A)}}_{\text{sinc}}
%\underbrace{\textcolor{orange}{\!\!\begin{bmatrix} 2 \\2 \\2\\2\end{bmatrix}\!\odot\!}}_{Q \text{in time}} \overbrace{\underbrace{\textcolor{blue}{\!\!\begin{bmatrix}
%    1\\ 0\\1\\0
%    \end{bmatrix}\!\odot\!}}_{\delta^{\scriptsize\textrm{(spl)}}} \underbrace{\!\!\begin{bmatrix}
%    s_0\\s_1\\s_2\\s_3
%    \end{bmatrix}}_{s}}^{s_\delta} \!\!\xrightarrow{\mathcal{F}} \!\! \underbrace{\textcolor{red}{\begin{bmatrix}
%   1\\1\\0\\0 \end{bmatrix}}}_{\text{l.p.f.}}\!\odot\! \
\underbrace{\textcolor{red}{\text{P}(A)}}_{\text{sinc}}\!\!
\underbrace{\textcolor{orange}{\!\!\begin{bmatrix} 2 \\ 0 \\ 2 \\ 0 \end{bmatrix}\!\odot\!}}_{\tiny Q \text{\tiny in time}}  \!\! \overbrace{\underbrace{\textcolor{blue}{\!\!\begin{bmatrix}
    1\\ 0\\1\\0
    \end{bmatrix}}}_{\delta^{\textrm{(spl)}}} \!\!\!\!\odot\!\! \underbrace{\!\!\begin{bmatrix}\!
    s_0\\s_1\\s_2\\s_3\!
    \end{bmatrix}}_{s}}^{s_\delta} \!\!\!\!\xrightarrow{\mathcal{F}} \!\!\!\!\\ \underbrace{\textcolor{red}{\!\!\begin{bmatrix}
   1\\1\\0\\0 \end{bmatrix}}}_{\text{l.p.f.}} \!\!\!\!\!\odot \, \underbrace{\textcolor{orange}{2I_4}}_{Q} \overbrace{\underbrace{\textcolor{blue}{\frac{1}{2}
    \begin{bmatrix}
    I_2 & I_2 \\ I_2 & I_2
    \end{bmatrix}}}_{\text{P}(M)}
    \underbrace{\begin{bmatrix}
    \widehat{s_0}\\ \widehat{s_1}\\0\\0
    \end{bmatrix}}_{\widehat{s}}}^{\widehat{s_\delta}}
\end{equation}%%
%}}%
%
By moving the factor of 2 in \eqref{recoveryq1} from $Q$ to the low-pass filter, and similarly moving the factor of 2 from $Q$ in time into the sinc, we achieve the traditional Nyquist-Shannon sampling recovery in~\eqref{eqn:recoverynyquist}. Equation~\eqref{eqn:recoverynyquist} is a simplified version of \eqref{recoveryq1}, removing the filter~$Q$ since it is the identity.
    \item $Q_2 = \begin{bmatrix} I_2 & I_2\\ \end{bmatrix}$. Choose the remaining $N-K$ rows of $Q$ to produce LSI filter $Q=1\cdot1^T\otimes I_2$, a matrix with four $I_2$ blocks. This~$Q$ is equivalent to multiplying $s_\delta$ by $[2,0,2,0]^T = 2\delta^{\scriptsize\textrm{(spl)}}$ in the time domain.
     %\widetilde{}
 %    \begin{center}
%
%     {{{\scriptsize
\vspace*{-.2cm}
\begin{equation} \label{recoveryq2}
\underbrace{\textcolor{red}{\text{P}(A)}}_{\text{sinc}}\!\!
\underbrace{\textcolor{orange}{\!\!\begin{bmatrix} 2 \\ 0 \\ 2 \\ 0 \end{bmatrix}\!\odot\!}}_{\tiny Q \text{\tiny in time}}  \!\! \overbrace{\underbrace{\textcolor{blue}{\!\!\begin{bmatrix}
    1\\ 0\\1\\0
    \end{bmatrix}}}_{\delta^{\textrm{(spl)}}} \!\!\!\!\odot\!\! \underbrace{\!\!\begin{bmatrix}\!
    s_0\\s_1\\s_2\\s_3\!
    \end{bmatrix}}_{s}}^{s_\delta} \!\!\!\!\xrightarrow{\mathcal{F}} \!\!\!\!\\ \underbrace{\textcolor{red}{\!\!\begin{bmatrix}
   1\\1\\0\\0 \end{bmatrix}}}_{\text{l.p.f.}} \!\!\!\!\!\odot \,
   \underbrace{ \textcolor{orange}{\!\!\begin{bmatrix}\!
    I_2 & I_2 \\ I_2 & I_2\!
    \end{bmatrix}}}_{Q} \overbrace{\underbrace{\textcolor{blue}{\frac{1}{2}
    \!\!\begin{bmatrix}
    I_2 & I_2 \\ I_2 & I_2
    \end{bmatrix}}}_{\text{P}(M)}\!\!
    \underbrace{\!\!\begin{bmatrix}
    \widehat{s_0}\\ \widehat{s_1}\\0\\0
    \end{bmatrix}}_{\widehat{s}}}^{\widehat{s_\delta}}
\end{equation}%%
%}}}
%\end{center}
Since $Q P(M) = 2 P(M)$, move the factor of 2 from $Q$ in time into the sinc and move the 2 in $2P(M)$ into the low-pass filter in~\eqref{recoveryq2}. By doing so, achieve traditional Nyquist-Shannon sampling recovery in~\eqref{eqn:recoverynyquist}. Equation~\eqref{eqn:recoverynyquist} is a simplified version of \eqref{recoveryq2} by replacing $Q P(M)$ with $2 P(M)$.
    \item $Q_2 = 2 \begin{bmatrix} 1 & 0 & 0 & 0\\ 0 & 0 & 0 & 1 \end{bmatrix}$. Now $Q$ cannot be LSI since the main diagonal must be constant. Filter $Q$ is equivalent to multiplying by a filter $\text{DFT}^{-1} Q \text{DFT}$ in the time domain.% by \eqref{eqn:genfilttime}.
Since $Q$ is not LSI, the time domain filter is not diagonal and not pointwise multiplication.
%{{\scriptsize
\begin{equation} \label{recoveryq3}
\underbrace{\textcolor{red}{\text{P}(A)}}_{\text{sinc}}
\underbrace{\textcolor{orange}{\text{DFT}^{-1} Q \text{DFT}}}_{Q \text{ in time}} \!\odot \! \overbrace{\underbrace{\textcolor{blue}{\!\begin{bmatrix}
    1\\ 0\\1\\0
    \end{bmatrix}\! \odot\!}}_{\delta}\! \underbrace{\!\begin{bmatrix}
    s_0\\s_1\\s_2\\s_3
    \end{bmatrix}}_{s}}^{s_\delta} \!\!\!\xrightarrow{\mathcal{F}}\!\!\!\! \\ \underbrace{\textcolor{red}{\begin{bmatrix}
   1\\1\\0\\0 \end{bmatrix}\!\odot\!}}_{\text{l.p.f.}}  \textcolor{orange}{Q} \overbrace{\underbrace{\textcolor{blue}{\frac{1}{2}
    \begin{bmatrix}
    I_2 & I_2 \\ I_2 & I_2
    \end{bmatrix}}}_{\text{P}(M)}
    \underbrace{\!\!\!\!\begin{bmatrix}
    \widehat{s_0}\\ \widehat{s_1}\\0\\0
    \end{bmatrix}}_{\widehat{s}}}^{\widehat{x_\delta}}
\end{equation}%%
%}}
\end{s_enumerate-0}
This example looked at three different possibilities for $Q_2$. The first two choices for the upper part of~$Q$, $Q_2$, lead to LSI $Q$ and equation~\eqref{eqn:recoverynyquist} leads to Nyquist-Shannon recovery. The third $Q_2$ does not lead to an LSI filter, but can be used to recover the signal.

\begin{figure}[hbt]
    \centering
   \includegraphics[height=3.5cm, width=5cm]{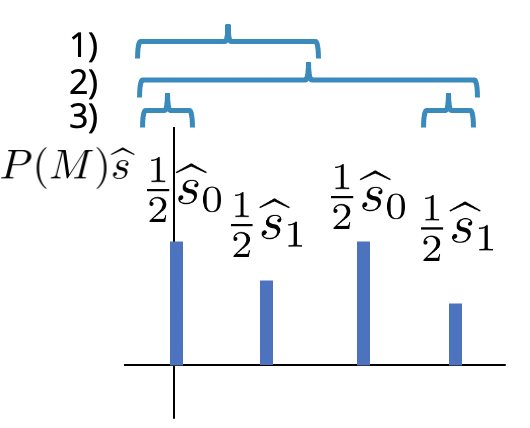}
    \caption{\small Values used by $Q_2$ to recover: values shown for each $Q_2$.}
    \label{fig:dspqkfig}
\end{figure}

All $Q_2$ use some values of $\widehat{s}_\delta$ to recover the signal. The first $Q_2$ uses the first and second values. The second $Q_2$ uses all four values. The third $Q_2$ uses the first and fourth values. This is shown in figure~\ref{fig:dspqkfig}.
Also, one might consider using the pseudoinverse of $P(M)_2$ as in \cite{gensample}. The pseudoinverse of $P(M)_2$ in~\eqref{conditionexdsp} is the second $Q_2$ considered above. It does indeed solve \eqref{conditionexdsp}, but it is only one of possible choices as illustrated by the theory above.

In DSP Nyquist-Shannon Sampling, $Q$ can be chosen as $\frac{N}{K} I_N$. The upper block of~$Q$ is $Q_K=I_K$ followed by $0_K$ matrices, scaled by $\frac{N}{K}$, as shown in~\eqref{recoveryq1}. This factor can be merged into the lowpass filter and $Q$ removed because it is the identity matrix. Thus, Nyquist-Shannon sampling recovery is a special, simplified case, where the $Q$ is the identity and is removed. However, the example shows that GSP allows for other upper blocks of~$Q$, $Q_K$, including non-LSI ones that recover~$s$ and that are not considered by traditional Nyquist-Shannon sampling.

\subsection{Connection to Frequency Domain Sampling in \cite{tanaka-2018, gensample}}
%\vspace*{-.15cm}
The frequency domain sampling proposed in \cite{tanaka-2018, gensample} does not correspond to the traditional concept of sampling as we explain now using our spectral filtering approach. In \cite{tanaka-2018}, let $U_0^*$ be the $\textrm{GFT}$ for the original undirected graph and $U_1$ be $\textrm{GFT}^{-1}$ for the sampled graph. Bandlimited graph signal~$s$ with band~$K$ is sampled in the frequency domain by:
	\begin{align}\label{eqn:Tanakasampling}
	\hspace{-.3cm}f = U_1 Q_1
	U_0^* s, \textrm{  with  } Q_1=\begin{bmatrix}
	\textrm{I}_{N/K} & \textrm{I}_{N/K} & \hdots & \textrm{I}_{N/K}
	\end{bmatrix}
	\end{align}
	where~$f$ is the sampled signal. Matrix $Q_1$  produces~$f$ whose $\textrm{GFT}$ spectrum is replicated like in DSP Nyquist-Shannon sampling. Matrix $Q_1$ is a decimated version of filter $Q=\left[Q_1^T\cdots Q_1^T\right]^T$ with $K\times K$ blocks.
Filtering with~$Q$ in the frequency domain multiplies by $\textrm{GFT}^{-1} Q \textrm{GFT}$ in the vertex domain. Since~$Q$ is circular, real, and symmetric $Q = \textrm{DFT}^H\cdot  \Lambda^*\cdot  \textrm{DFT}=\textrm{DFT}\cdot  \Lambda\cdot  \textrm{DFT}^H$. This is equivalent to multiplying in the vertex domain by $\widetilde{Q}=\textrm{GFT}^{-1}\cdot  \textrm{DFT}\cdot  \Lambda \cdot \textrm{DFT}^H \cdot \textrm{GFT}$. In general $\textrm{GFT}^{-1}\neq \textrm{DFT}^*$ and $\textrm{GFT}^{-1}\cdot\textrm{DFT}\neq I$. So, $\widetilde{Q}$ is not diagonal, and it cannot subsample in the vertex domain (keep some and discard other signal values). Because of this, filtering in the spectral domain with the approach in \cite{tanaka-2018} requires knowledge and distorts \textbf{all} of the signal' samples in the vertex domain. This is different from recoverability in traditional sampling where only a decimated signal is used.
%\color{black}

%\vspace*{-.4cm}
	\section{Conclusion} \label{sec:conclusion}
%	\vspace*{-.1cm}
	The literature on graph sampling is quite robust. Several approaches address the design of the sampling set~$S$ and develop alternative recovery methods. Some are developed in the vertex domain, others in the spectral domain. But the vertex and spectral domain sampling methods are not related lacking the dualism in DSP sampling. The goal of this paper is to present a unifying theory for GSP sampling showing the analogy and dualism between the vertex and spectral domain versions of all standard sampling steps. We introduce a graph spectral shift~$M$ and develop a spectral graph signal processing theory that is the dual of the vertex based GSP. We then show that GSP vertex subsampling is LSI filtering in the spectral domain with polynomials $P(M)$, decimation replicates the spectrum of the decimated signal, and interpolation is achieved by filtering operations in both vertex and spectral domains.  Examples illustrate the impact of choices that can be made in GSP and show how GSP sampling becomes DSP sampling when the graph is the directed cycle time graph.
	\vspace*{-.6cm}
	\bibliographystyle{ieeetr}
	\bibliography{refs,sampling}

\begin{thebibliography}{10}

\bibitem{Sandryhaila:13}
A.~Sandryhaila and J.~M.~F. Moura, ``Discrete signal processing on graphs,''
  {\em IEEE Trans. Signal Proc.}, vol.~61, pp.~1644--1656, April 2013.

\bibitem{ShumanNFOV:13}
D.~I. Shuman, S.~K. Narang, P.~Frossard, A.~Ortega, and P.~Vandergheynst, ``The
  emerging field of signal processing on graphs: {E}xtending high-dimensional
  data analysis to networks and other irregular domains,'' {\em IEEE Signal
  Proc. Magazine}, vol.~30, pp.~83--98, May 2013.

\bibitem{Sandryhaila:14}
A.~Sandryhaila and J.~M.~F. Moura, ``Discrete signal processing on graphs:
  Frequency analysis,'' {\em IEEE Trans. Signal Proc.}, vol.~62,
  pp.~3042--3054, June 2014.

\bibitem{Sandryhaila:14big}
A.~Sandryhaila and J.~M.~F. Moura, ``Big data analysis with signal processing
  on graphs: Representation and processing of massive data sets with irregular
  structure,'' {\em IEEE Signal Processing Magazine}, vol.~31, pp.~80--90,
  September 2014.

\bibitem{EldarTanakaSPM}
Y.~Tanaka, Y.~C. Eldar, A.~Ortega, and G.~Cheung, ``Sampling signals on graphs:
  From theory to applications,'' {\em IEEE Signal Processing Magazine},
  vol.~37, no.~6, pp.~14--30, 2020.

\bibitem{oppenheimwillsky-1983}
A.~V. Oppenheim and A.~S. Willsky, {\em Signals and Systems}.
\newblock Englewood Cliffs, New Jersey: Prentice-Hall, 1983.

\bibitem{chenvarmasinghkovacevic}
S.~{Chen}, R.~{Varma}, A.~{Singh}, and J.~{Kovacevi\'c}, ``Signal recovery on
  graphs: Fundamental limits of sampling strategies,'' {\em IEEE Tr. on Sig.
  and Inform. Proc. over Networks}, vol.~2, no.~4, pp.~539--554, 2016.

\bibitem{pesenson2008sampling}
I.~Pesenson, ``Sampling in {P}aley-{W}iener spaces on combinatorial graphs,''
  {\em Transactions of the American Mathematical Society}, vol.~360, no.~10,
  pp.~5603--5627, 2008.

\bibitem{pesenson2010sampling}
I.~Z. Pesenson and M.~Z. Pesenson, ``Sampling, filtering and sparse
  approximations on combinatorial graphs,'' {\em Journal of Fourier Analysis
  and Appl.}, vol.~16, no.~6, pp.~921--942, 2010.

\bibitem{Narang:12}
S.~K. Narang and A.~Ortega, ``Perfect reconstruction two-channel wavelet filter
  banks for graph structured data,'' {\em IEEE Trans. Signal Proc.}, vol.~60,
  no.~6, pp.~2786--2799, 2012.

\bibitem{anis2014towards}
A.~Anis, A.~Gadde, and A.~Ortega, ``Towards a sampling theorem for signals on
  arbitrary graphs,'' in {\em 2014 IEEE Int. Conf. on Acoustics, Speech and
  Signal Processing (ICASSP)}, pp.~3864--3868, IEEE, 2014.

\bibitem{gadde2015probabilistic}
A.~Gadde and A.~Ortega, ``A probabilistic interpretation of sampling theory of
  graph signals,'' in {\em 2015 IEEE Int. Conf. on Acoustics, Speech and Signal
  Processing (ICASSP)}, pp.~3257--3261, IEEE, 2015.

\bibitem{anis2016efficient}
A.~Anis, A.~Gadde, and A.~Ortega, ``Efficient sampling set selection for
  bandlimited graph signals using graph spectral proxies,'' {\em IEEE Trans.
  Signal Proc.}, vol.~64, no.~14, pp.~3775--3789, 2016.

\bibitem{anis2017critical}
A.~Anis and A.~Ortega, ``Critical sampling for wavelet filterbanks on arbitrary
  graphs,'' in {\em 2017 IEEE International Conference on Acoustics, Speech and
  Signal Processing (ICASSP)}, pp.~3889--3893, IEEE, 2017.

\bibitem{Jelena}
S.~Chen, R.~Varma, A.~Sandryhaila, and J.~Kova\v{c}evi\'c, ``Discrete signal
  processing on graphs: Sampling theory,'' {\em IEEE Trans. Signal Proc.},
  vol.~63, no.~24, pp.~6510 -- 6523, 2015.

\bibitem{tremblay2017graph}
N.~Tremblay, P.-O. Amblard, and S.~Barthelm{\'e}, ``Graph sampling with
  determinantal processes,'' in {\em 2017 25th European Signal Proc.~Conf.
  (EUSIPCO)}, pp.~1674--1678, IEEE, 2017.

\bibitem{tsitsvero2016signals}
M.~Tsitsvero, S.~Barbarossa, and P.~Di~Lorenzo, ``Signals on graphs:
  Uncertainty principle and sampling,'' {\em IEEE Transactions on Signal
  Processing}, vol.~64, no.~18, pp.~4845--4860, 2016.

\bibitem{marques2015sampling}
A.~G. Marques, S.~Segarra, G.~Leus, and A.~Ribeiro, ``Sampling of graph signals
  with successive local aggregations,'' {\em IEEE Trans.~on Signal Processing},
  vol.~64, no.~7, pp.~1832--1843, 2015.

\bibitem{jungheromarajahromiheimowitzeldar-2019}
A.~{Jung}, A.~O. {Hero, III}, A.~C. {Mara}, S.~{Jahromi}, A.~{Heimowitz}, and
  Y.~C. {Eldar}, ``Semi-supervised learning in network-structured data via
  total variation minimization,'' {\em IEEE Trans.~on Signal Processing},
  vol.~67, no.~24, pp.~6256--6269, 2019.

\bibitem{chen2015signal-2}
S.~Chen, A.~Sandryhaila, J.~M.~F. Moura, and J.~Kova{\v{c}}evi{\'c}, ``Signal
  recovery on graphs: Variation minimization,'' {\em IEEE Transactions on
  Signal Processing}, vol.~63, no.~17, pp.~4609--4624, 2015.

\bibitem{eldar-samplingbook2015}
Y.~C. Eldar, {\em Sampling Theory: Beyond Bandlimited Systems}.
\newblock USA: Cambridge University Press, 1st~ed., 2015.

\bibitem{gensample}
Y.~{Tanaka} and Y.~{Eldar}, ``Generalized sampling on graphs with subspace and
  smoothness priors,'' {\em IEEE Tr. on Signal Proc.}, vol.~68, 2020.

\bibitem{tanaka-2018}
Y.~{Tanaka}, ``Spectral domain sampling of graph signals,'' {\em IEEE
  Transactions on Signal Processing}, vol.~66, no.~14, pp.~3752--3767, 2018.

\bibitem{ortegafrossardkovacevicmouravandergheynst-2018}
A.~{Ortega}, P.~{Frossard}, J.~{Kovačević}, J.~M.~F. {Moura}, and
  P.~{Vandergheynst}, ``Graph signal processing: Overview, challenges, and
  applications,'' {\em Proceedings of the IEEE}, vol.~106, pp.~808--828, May
  2018.

\bibitem{giraultgoncalvesfleury-2015}
B.~Girault, P.~Gonçalves, and {\'E}.~Fleury, ``Translation on graphs: An
  isometric shift operator,'' {\em IEEE Signal Processing Letters}, vol.~22,
  pp.~2416--2420, Dec 2015.

\bibitem{gavilizhang-2017}
A.~Gavili and X.~P. Zhang, ``On the shift operator, graph frequency, and
  optimal filtering in graph signal processing,'' {\em IEEE Tr.~on Signal
  Proc.}, vol.~65, pp.~6303--6318, Dec 2017.

\bibitem{derimoura-2017}
J.~A. Deri and J.~M.~F. Moura, ``Spectral projector-based graph {F}ourier
  transforms,'' {\em IEEE Journal of Selected Topics in Signal Processing},
  vol.~11, pp.~785--795, Sept 2017.

\bibitem{shimoura-asilomar2019}
J.~Shi and J.~M.~F. Moura, ``Topics in graph signal processing: Convolution and
  modulation,'' in {\em {(ACSSC)} Asilomar Conference on Signals, Systems, and
  Computers}, IEEE, 2019.

\bibitem{shi2019graph}
J.~Shi and J.~M.~F. Moura, ``Graph signal processing: Modulation, convolution,
  and sampling,'' December 2019.

\bibitem{leus2017dual}
G.~Leus, S.~Segarra, A.~Ribeiro, and A.~G. Marques, ``The dual graph shift
  operator: Identifying the support of the frequency domain,'' 2017.

\bibitem{jayant1982analog}
N.~S. Jayant, ``Analog scramblers for speech privacy,'' {\em Computers \&
  Security}, vol.~1, no.~3, pp.~275--289, 1982.

\bibitem{SakuraiKogaMuratani-1984}
K.~Sakurai, K.~Koga, and T.~Muratani, ``A speech scrambler using the fast
  {F}ourier transform technique,'' {\em IEEE Journal on Selected Areas in
  Comm.}, vol.~2, no.~3, pp.~434--442, 1984.

\bibitem{pesenson2001sampling}
I.~Pesenson, ``Sampling of band-limited vectors,'' {\em Journal of {F}ourier
  Analysis and Appl.}, vol.~7, no.~1, pp.~93--100, 2001.

\bibitem{Vaidyanathan}
P.~Vaidyanathan, {\em Multirate Systems and Filter Banks}.
\newblock Prentice Hall, 1993.

\end{thebibliography}
	\vspace{-1cm}
		\begin{IEEEbiography}[{\includegraphics[width=1.1in,height=1.2in,clip,keepaspectratio]{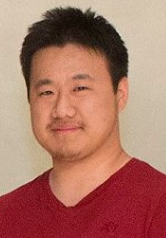}}]{John Shi} received his B.S. degrees in Computer Engineering and Applied Mathematics in 2017 from the University of Maryland, College Park.
		He is currently a Ph.D. student in the Electrical and Computer Engineering Department at Carnegie Mellon University. His research interests include graph signal processing theory, applications, and graph convolutional neural networks.
\end{IEEEbiography}
	\vspace*{-1.5cm}
	\begin{IEEEbiography}[{\includegraphics[width=1in,height=1.25in,clip,keepaspectratio]{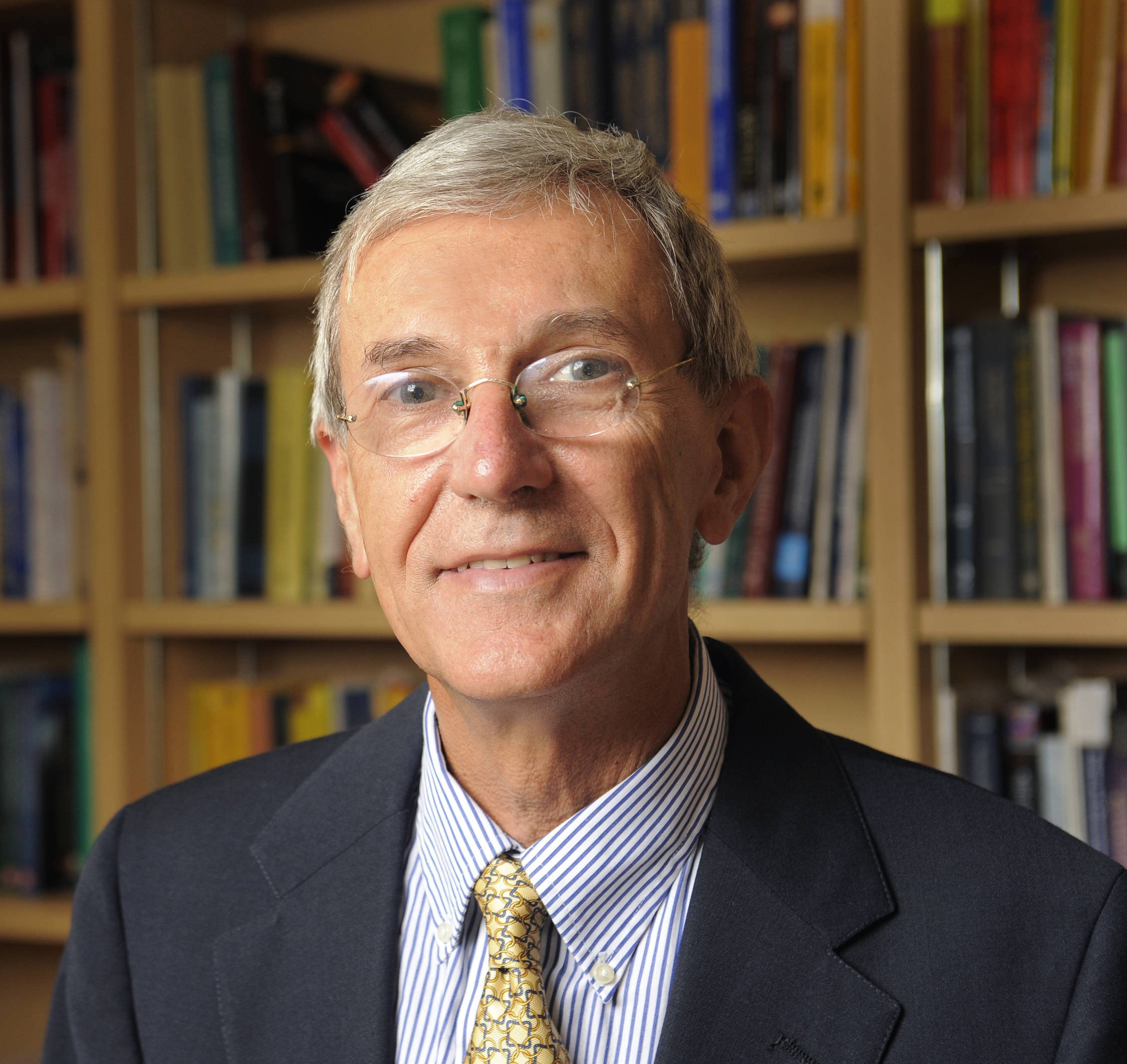}}]{Jos\'e M.~F.~Moura}(S'71--M'75--SM'90--F'94) is the Philip L.~and Marsha Dowd University Professor at Carnegie Mellon University (CMU). He holds degrees from Instituto Superior T\'ecnico (IST), Lisbon, Portugal, and  from the Massachusetts Institute of Technology (MIT), Cambridge, MA. He was on the faculty at IST and has held visiting faculty appointments at MIT and New York University (NYU). He founded and directs a large education and research program between CMU and Portugal, www.cmuportugal.org.

His research interests are on  data science and graph signal processing. Two of his patents (co-inventor A. Kav\v{c}i\'c) are used in over four billion disk drives in 60~\% of all computers sold in the last 15 years worldwide and were, in 2016, the subject of a 750 million dollar settlement between CMU and a chip manufacturer, the largest ever university verdict/settlement in the information technologies area.

Dr. Moura was the 2019 IEEE President and CEO. He has been Editor in Chief for the IEEE Transactions in Signal Processing.

Dr. Moura received the Technical Achievement Award and the Society Award from the IEEE Signal Processing Society, and the CMU College of Engineering Distinguished Professor of Engineering Award. He is a Fellow of the IEEE, the American Association for the Advancement of Science (AAAS), a corresponding member of the Academy of Sciences of Portugal, Fellow of the US National Academy of Inventors, and a member of the US National Academy of Engineering. He received Doctor Honoris Causa degrees from the University of Strathclyde and Universidade de Lisboa. He was awarded the  Gr\~a Cruz Infante D. Henrique, bestowed to him by the President of the Republic of Portugal.
\end{IEEEbiography}
\end{document}